\PassOptionsToPackage{table}{xcolor}
\pdfoutput=1

\documentclass[11pt]{article}
\usepackage{xcolor}
\usepackage{EMNLP2023}

\usepackage{times}
\usepackage{latexsym}

\usepackage[T1]{fontenc}

\usepackage[utf8]{inputenc}

\usepackage{microtype}

\usepackage{inconsolata}

\usepackage{amsmath}         
\usepackage{enumitem}        
\usepackage{graphicx}        
\usepackage{multirow}        
\usepackage{booktabs}        

\usepackage{float}
\usepackage{listings}
\usepackage{nicefrac}

\usepackage{tcolorbox}
\usepackage{tikz}
\usepackage[font=small]{caption}
\usetikzlibrary{matrix, positioning}
\usepackage{xspace}
\usepackage{subcaption}  
\usepackage{amssymb}    
\usepackage{mathrsfs}   
\usepackage{eucal}      
\usepackage{amsfonts}   

\renewcommand{\arraystretch}{1.1} 
\usepackage{pifont} 
\usepackage{newunicodechar}
\newunicodechar{✓}{\ding{51}}
\newunicodechar{✗}{\ding{55}}

\usepackage{array} 
\usepackage{makecell}
\newenvironment{packeditemize}{
\begin{itemize}[leftmargin=*]
  \setlength{\itemsep}{0pt}
  \setlength{\parskip}{0pt}
  \setlength{\parsep}{0pt}
}{\end{itemize}}

\title{\systemname: Synthesizing Relational Databases from Unstructured Text}

\author{
  \textbf{Mushtari Sadia} \quad
  \textbf{Zhenning Yang} \quad
  \textbf{Yunming Xiao} \\
  \textbf{Ang Chen} \quad
  \textbf{Amrita Roy Chowdhury} \\
  University of Michigan, Ann Arbor \\
  \texttt{\{mushtari, znyang, yunmingx, chenang, aroyc\}@umich.edu}
}

\begin{document}
\newcommand{\arc}[1]{\textcolor{purple}{ARC: #1}}
\newcommand{\yx}[1]{\textcolor{blue}{YX: #1}}
\newcommand{\ms}[1]{\textcolor{brown}{MS: #1}}
\newcommand{\systemname}{SQUiD\xspace}
\newcommand{\taskname}{Text2R}

\newcommand{\mOne}{$\mathbb{T}$\xspace}
\newcommand{\mTwo}{$\mathbb{S}$\xspace}
\newcommand{\mThree}{$\mathbb{L}$\xspace}

\newcommand{\eOne}{$\mathbb{T}\mathbin{\oplus}$\mTwo\xspace}
\newcommand{\eTwo}{$\mathbb{T}\mathbin{\oplus}$\mThree\xspace}
\newcommand{\eThree}{\systemname\xspace}

\newcommand{\github}{\url{https://anonymous.4open.science/r/SQUiD-79F4/}}

\maketitle

\begin{abstract}
Relational databases are central to modern data management, yet most data exists in unstructured forms like text documents. To bridge this gap, we leverage large language models (LLMs) to automatically synthesize a relational database by generating its schema and populating its tables from raw text. We introduce \systemname, a novel neurosymbolic framework that decomposes this task into four stages, each with specialized techniques. Our experiments show that \systemname consistently outperforms baselines across diverse datasets. Our code and datasets are publicly available at: \href{https://github.com/Mushtari-Sadia/SQUiD}{https://github.com/Mushtari-Sadia/SQUiD}.

\end{abstract}

\section{Introduction}

Relational databases serve as the foundation for data management, supported by decades of mature infrastructure development and a wide array of sophisticated analytical tools. 
However, much of today's data exists as raw, unstructured text -- such as academic articles, medical records, and business reports~\cite{unstructured-data}. This unstructured data cannot be directly analyzed using conventional database tools, which rely on structured, relational inputs. Bridging this gap remains a long-standing goal of the data management community~\cite{Mansuri-uns-to-db,dmg1,dmg2,dmg3,dmg4,dmg5,dmg6}, with a key challenge being the conversion of unstructured text into queryable, structured formats compatible with existing relational database infrastructure.

Large language models (LLMs) presents a unique opportunity to \textit{automate} this conversion, owing to their growing capability to understand natural language and perform complex information extraction tasks.   Prior work in this space can be broadly categorized into two areas. The first focuses on generating summarizing structures from text, such as tables \cite{deng2024texttupletableinformationintegrationtexttotable, wu2022texttotable, sundar2024gtbls, li2023seq2seqset,Aroravldb} and mind maps \cite{jain2024structsumgenerationfastertext}---but these non-relational representations are often tailored for specific downstream applications~\cite{shavarani2024entityretrievalansweringentitycentric, 10.1145/3616855.3635752}, and lack the expressiveness and semantics of relational databases. The second category manipulates a \textit{pre-defined} and fully populated relational database---e.g., Text-to-SQL~\cite{hong2024next} approaches generate executable SQL queries from text over given schemas, while a recent work can update existing relational databases using text input \cite{jiao-etal-2024-text2db}. However, a key challenge of managing  unstructured text is precisely that such a pre-defined database often does not exist.

\begin{figure}[t]
    \centering
    \includegraphics[width=0.9\linewidth]{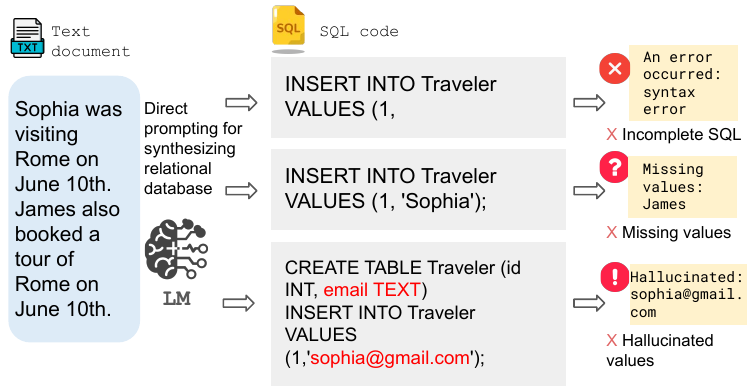}\vspace{-0.2cm}
    \caption{Challenges of synthesizing relational DB from text}
    \label{fig:dbm-failure}
    \vspace{-6mm}
\end{figure}
In this paper, we pursue a more ambitious goal -- \textit{synthesizing a relational database from unstructured text from scratch}---a task that we call \taskname. The \taskname~task presents 
several unique challenges. First, a relational schema consists of multiple interrelated tables that capture complex entity-relationship semantics, and it must also preserve syntactic integrity, such as satisfying primary/foreign key constraints.
\begin{figure*}
    \centering
    \includegraphics[width=0.99\linewidth]{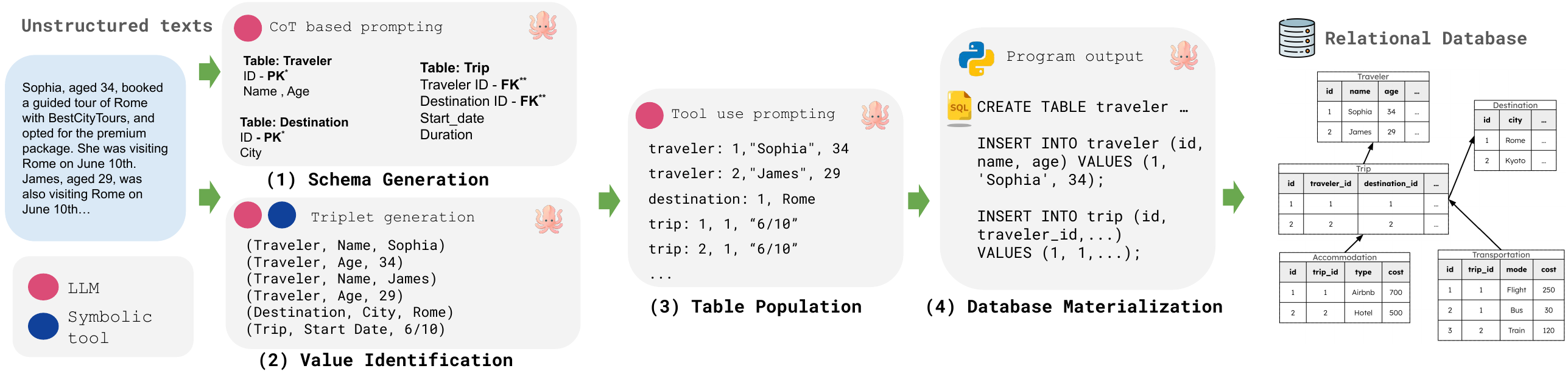}\vspace{-0.2cm}
    \caption{Overview of \systemname. \textbf{(1) Schema Generation} constructs a relational schema that defines the tables, columns, and their relationships, from the entities in the text. \textbf{(2) Value Identification} extracts relevant values (e.g., names, dates) from the text. These values are then organized during \textbf{(3) Table Population} by aligning them with the generated schema to form tuples. \textbf{(4) Database Materialization} programmatically translates the output into \texttt{SQL} statements, producing the final relational database.}
    \label{fig:e2e}
    \vspace{-5mm}
\end{figure*}
Second, database records must be correctly identified and populated across tables. This involves ensuring value consistency -- e.g., the same entity must be consistently represented in all relevant tables.
Third, the actual database creation requires valid and executable SQL statements, adding another layer of complexity. 
Na\"{i}ve approaches, such as directly prompting LLMs to synthesize databases, leads to diverse errors, including missing or hallucinated values, and SQL syntax issues (Fig. \ref{fig:dbm-failure}).

To address these challenges, we propose \systemname\footnote{SQUiD - \textbf{SQ}L on \textbf{U}nstructured \textbf{D}ata}, a neurosymbolic framework for the \taskname~task. 
Our key idea is to decompose the task into multiple modular stages in a principled manner---breaking the problem into manageable sub-tasks. This allows each stage to leverage specialized techniques, such as symbolic information extraction and LLM-assisted tool use, for improved performance. Via task breakdown, some stages can also be executed programmatically, enhancing both accuracy and consistency. Additionally, each stage incorporates best practices from relational database literature to guide prompt design.

\systemname consists of four stages, which generalize across text from diverse domains. 
The \textit{schema generation} stage uses LLMs to infer a relational schema from the input text, guided by carefully designed prompts that incorporate best practices to identify entities and relationships. In the \textit{value identification} stage, intermediate representations in the form of triplets are extracted using both symbolic tools and LLMs. These triplets break down complex sentences into granular units, improving coverage of the extracted values.
Next, the \textit{table population} stage aligns these triplets with the generated schema to form schema-consistent tuples.
Finally, instead of generating SQL directly via LLMs—which can be token-intensive—our \textit{database materialization} stage programmatically translates the structured outputs into valid SQL statements, ensuring syntactic correctness and structural fidelity. The resulting SQL is then executed to instantiate the final database.
We make the following contributions:
\vspace{-0.2cm}


\begin{packeditemize}

    \item We define a new task -- synthesizing relational databases from unstructured text, or \taskname. This marks a clear departure from prior work, which  focuses on downstream relational tasks (e.g., Text2SQL), assuming a pre-existing database. 

        \item We propose \systemname, a novel neurosymbolic framework for \taskname, based on a four-stage decomposition. Each stage leverages custom techniques tailored to its specific subtask.

    \item We establish an automated benchmark methodology for \taskname. We also define a suite of evaluation metrics to assess schema and tuple quality along both semantic and syntactic dimensions.

    \item 
   We conduct extensive experiments across diverse text domains and show that \systemname consistently outperforms direct prompting baselines. 
    
\end{packeditemize}
\vspace{-0.4cm}

\begin{figure*}
    \centering
    \includegraphics[width=1\linewidth]{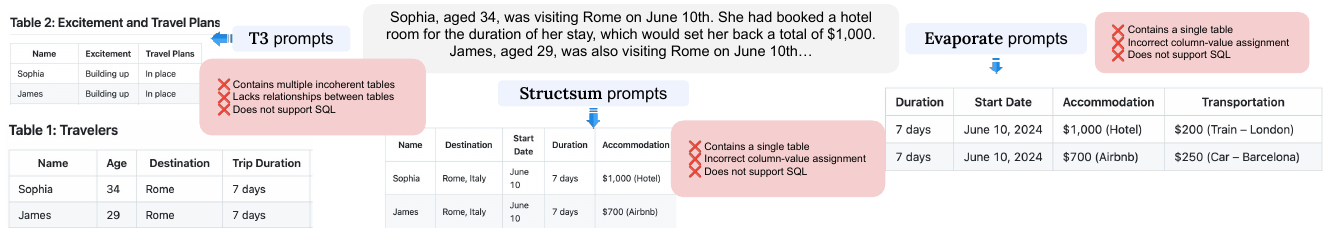}
\caption{Closest related works—\textsc{T3}\cite{deng2024texttupletableinformationintegrationtexttotable}, \textsc{StructSum}\cite{jain2024structsumgenerationfastertext}, and \textsc{Evaporate}\cite{Aroravldb}—when applied to our example dataset, either produced a single table with incorrect column-value assignments or multiple disconnected, irrelevant tables. In contrast, as shown in Fig.\ref{fig:e2e}, \systemname~correctly generates all five tables corresponding to the entities (\textit{Traveler}, \textit{Trip}, \textit{Accommodation}, \textit{Transportation} and \textit{Destination}) along with their proper relationships.} \vspace{-0.5cm}
    \label{fig:related-work}
\end{figure*}
\section{The \taskname~Task}\vspace{-0.2cm}
\label{sec:task}

We begin by defining this new task of relational database synthesis, or \taskname. Given an unstructured document \( D \) of natural language text, the goal is to produce a set of SQL statements \( S \): (1) \texttt{CREATE TABLE} statements which define the schema \( \mathcal{R} \), specifying the structure of the database in terms of tables and columns; and 
(2) \texttt{INSERT} statements which populate the 
relations with data extracted from the text in $D$.
The schema \( \mathcal{R} \) consists of a set of tables \(\textbf{T} =  \{T_1, T_2, \dots, T_n\} \) where each $T_i$ has a set of columns \( \textbf{C}_i = \{C_{i,1}, C_{i,2}, \dots, C_{i,k_i}\} \). Each table corresponds to an entity type, and the tables are inter-related,
organizing the extracted tuples from the text into a database. A tuple \( t \) for table \( T_i \) is represented as:
$t = \langle v_{1}, v_{2}, \dots, v_{k_i} \rangle$
where \( v_{j} \) is the value corresponding to column \( C_{ij} \in T_i \). Each tuple represents a unique \textit{instance} of the entity described by \( T_i \). Fig~\ref{fig:related-work} illustrates the differences between \taskname~and other tasks. 



\vspace{-0.2cm}

\section{\systemname\ Framework}
\vspace{-0.2cm}




\systemname~decomposes the \taskname~task into four modular stages that mirror the typical database construction process. First, a relational database schema is designed by identifying the domain's entities and relationships—this is the \textbf{schema generation} stage. Next, \systemname~extracts all the relevant values from the text (\textbf{value identification}), which are then used to construct tuples (\textbf{table population}). Finally, the generated schema and tuples are translated into valid SQL statements during the \textbf{database materialization} stage.
We describe these stages below, using the following text shown in Fig.~\ref{fig:e2e} as a running example: 
\textit{“Sophia booked a guided tour of Rome with BestCityTours, and opted for the premium package. She was visiting Rome on June 10th. James, aged 29, was also visiting Rome on June 10th.”}\\
\label{par:example}




\vspace{-0.5cm}\subsection{Schema Generation}\vspace{-0.1cm}
\label{subsec-schema-gen}

\begin{figure}[tbh]
\vspace{-1em}
    \centering
    \includegraphics[width=0.8\linewidth]{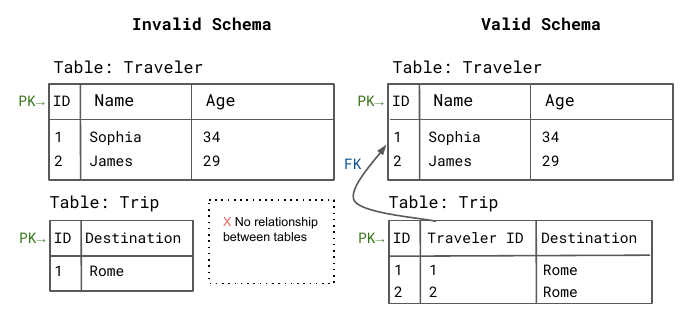} \vspace{-0.5cm}
    \caption{Examples of valid versus invalid relational schemas. PK: Primary key; FK: Foreign key.} 
    \label{fig:schema}\vspace{-1em}
\end{figure}

\textbf{Challenge.} The complexity of schema generation is both semantic and syntactic. Semantically, the schema must accurately capture the entity-relationship structure that reflects the underlying data. Syntactically, a valid schema must comply with the integrity constraints defined by the established principles of relational databases. Simply prompting LLMs to generate a schema without explicitly articulating the necessary relational database constraints can result in structurally invalid outputs, as illustrated in Fig.~\ref{fig:schema}. 
\\\noindent\textbf{Approach.} The novelty of our approach is to encode a standardized set of rules  that reflect the best practices in relational database literature, effectively guiding the model through a structured design process. These rules cover: (1) identifying relevant entities and relationships, (2) defining tables with appropriate columns, (3) assigning primary and foreign keys, and (4) avoiding reserved SQL keywords in naming tables/columns. We encode these rules into two types of prompt strategies: direct, and chain-of-thought (CoT) prompting. CoT decomposes schema generation into intermediate reasoning steps (e.g., entity identification, then table and key definition;  see Appendix~\ref{appendix:prompt-schema-cot}). 

Decoupling schema generation from tuple formation has another advantage -- it allows schema validity to be evaluated in isolation. This modularity is essential for enforcing syntactic constraints: each table must define a primary key (a column, or set of columns that uniquely identifies each row); and tables should include foreign keys (columns referencing primary keys in other tables). These constraints capture relationships between tables and enable \texttt{JOIN} operations. 


\vspace{-0.2cm}
\subsection{Value Identification}\vspace{-0.1cm}
\label{subsec-value-iden}

\textbf{Challenge.} This stage identifies and extracts values from the text that correspond to columns across all tables in the schema, presenting two challenges. 
First, multiple values often need to be extracted and deduplicated from the input to form a complete tuple (i.e., an entity instance). In our example, 
\textit{“Sophia booked a guided tour of Rome with BestCityTours, and opted for the premium package. She was visiting Rome on June 10th.”}, we must recover several values, such as traveler name ("Sophia"), tour location ("Rome"), tour operator ("BestCityTours"), and date ("June 10th"); redundant mentions (e.g. "Rome") need to be detected and deduplicated. Second, a document may describe multiple instances of the same type of entity, so we need to assign each value to the correct tuple. For instance, in the passage we also have:
\textit{“James, aged 29, was also visiting Rome on June 10th.”} Hence, we need to track that Sophia and James are different tourists, and form distinct tuples. 

\noindent\textbf{Approach.} 
Our neurosymbolic approach first augments direct LLM prompting with two information extraction (IE) methods to isolate values in a structured format, and then guides the LLM to accurately group these values by tuples. 


\noindent\textit{Triplet Generation.} This step introduces an intermediate representation using \textit{triplets}, a format commonly used in information extraction. 
Specifically, we consider two triplet formats:
\vspace{-0.2cm}\begin{packeditemize}
\item \textbf{Symbolic triplets}, in the form \texttt{(subject}, \texttt{relation}, \texttt{object)}—e.g.,
\texttt{(Sophia}, \texttt{visiting}, \texttt{Rome)},
extracted symbolically using the Stanford CoreNLP toolkit  \cite{manning-etal-2014-stanford}.
\item \textbf{Schema-aligned triplets}, in the form \texttt{(table} \texttt{column}, \texttt{value)}—e.g.,
\texttt{(Tour}, \texttt{Location}, \texttt{Rome)},
generated using prompt-based LLM extraction for the target schema (see Appendix~\ref{appendix:prompts}).
\end{packeditemize}
\vspace{-0.2cm}
For instance, the earlier passage describing Sophia might yield the following schema-aligned triplets:

\tikzset{every picture/.style={line width=0.75pt}} 

\begin{tikzpicture}[x=0.75pt,y=0.75pt,yscale=-1,xscale=1]

\draw   (36,23) -- (161,23) -- (161,60) -- (36,60) -- cycle ;
\draw   (167,23) -- (279,23) -- (279,60) -- (167,60) -- cycle ;

\draw (39,27) node [anchor=north west][inner sep=0.75pt]  [font=\tiny] [align=left] {<Traveler, Name, Sophia>};
\draw (39,38) node [anchor=north west][inner sep=0.75pt]  [font=\tiny] [align=left] {<Trip, Destination, Rome>};
\draw (39,50) node [anchor=north west][inner sep=0.75pt]  [font=\tiny] [align=left] {<Booking, Date, June 10th>};
\draw (70,10) node [anchor=north west][inner sep=0.75pt]   [align=left] {\textbf{{\scriptsize Sophia}}};
\draw (170,27) node [anchor=north west][inner sep=0.75pt]  [font=\tiny] [align=left] {<Traveler, Name, James>};
\draw (170,38) node [anchor=north west][inner sep=0.75pt]  [font=\tiny] [align=left] {<Trip, Destination, Rome>};
\draw (170,50) node [anchor=north west][inner sep=0.75pt]  [font=\tiny] [align=left] {<Booking, Date, June 10th>};
\draw (200,10) node [anchor=north west][inner sep=0.75pt]  [font=\scriptsize] [align=left] {\textbf{James}};

\end{tikzpicture}

We consider these two types of triplets because each captures complementary sets of values. Symbolic tools use deterministic methods to parse the text, and often extract values that LLMs may overlook (e.g. modifier words like \textit{premium}). In contrast, LLM-generated schema-aligned triplets are more structurally consistent with the database schema, (e.g., \textit{Location $\rightarrow$ Rome}). 

To ensure comprehensive coverage, we additionally leverage part-of-speech (POS) tagging to identify all nouns, pronouns, and numerical tokens in the text, since these POS categories typically encompass most values. We then perform string matching to verify whether the extracted triplets include all such tokens. If any are missing, the LLM is prompted to augment the existing triplets by incorporating the missing POS tokens. 
\\
\noindent\textit{Triplet Deduplication.}  Both triplet generation methods often introduce redundancy. To reduce this, we use the "sentence-t5-base" model \cite{ni2021sentencet5scalablesentenceencoders} to generate embeddings of the triplets and apply cosine similarity to identify near-duplicates. If a set of triplets has a pairwise cosine similarity above a tunable threshold (97\%), we retain only one representative triplet.

\noindent\textit{Triplet Grouping.} 
To ensure that triplets are correctly grouped by entity instance, we apply two heuristics. First, we assume that the first table in the schema typically corresponds to the central entity (e.g., the tourist in a tourism booking system). Second, we leverage the structure of the input document, where each paragraph often describes a distinct instance of this central entity. Accordingly, we associate each paragraph with a unique \textit{identifier}, which serves as the primary key for the first table. In particular, \systemname~uses an LLM to detect the number of distinct entity instances in the document and assign a unique identifier to each paragraph. Once assigned, each triplet is prefixed with its corresponding identifier. For example:

\tikzset{every picture/.style={line width=0.75pt}} 

\begin{tikzpicture}[x=0.75pt,y=0.75pt,yscale=-1,xscale=1]

\draw   (36,23) -- (166,23) -- (166,60) -- (36,60) -- cycle ;
\draw   (166,23) -- (286,23) -- (286,60) -- (166,60) -- cycle ;

\draw (39,27) node [anchor=north west][inner sep=0.75pt]  [font=\tiny] [align=left] {<1, Traveler, Name, Sophia>};
\draw (39,38) node [anchor=north west][inner sep=0.75pt]  [font=\tiny] [align=left] {<1, Trip, Destination, Rome>};
\draw (39,50) node [anchor=north west][inner sep=0.75pt]  [font=\tiny] [align=left] {<1, Booking, Date, June 10th>};
\draw (69,10) node [anchor=north west][inner sep=0.75pt]   [align=left] {\textbf{{\scriptsize Sophia}}};
\draw (170,27) node [anchor=north west][inner sep=0.75pt]  [font=\tiny] [align=left] {<2, Traveler, Name, James>};
\draw (170,38) node [anchor=north west][inner sep=0.75pt]  [font=\tiny] [align=left] {<2, Trip, Destination, Rome>};
\draw (170,50) node [anchor=north west][inner sep=0.75pt]  [font=\tiny] [align=left] {<2, Booking, Date, June 10th>};
\draw (200,10) node [anchor=north west][inner sep=0.75pt]  [font=\scriptsize] [align=left] {\textbf{James}};

\end{tikzpicture}

This structure ensures that all extracted values are correctly grouped by the entity instance they describe, and that the same identifier can be used to link rows across tables during the population stage.




\vspace{-0.2cm}
\subsection{Table Population}\vspace{-0.1cm}
\label{subsec-table-pop}

\textbf{Challenge.} This stage constructs tuples for each table using the values identified in the previous stage, presenting two challenges. First, each value must be correctly \textit{aligned} with its corresponding table column, meaning the LLM must output tuples in a schema-aligned format. However, extracting structured information in a single generation often results in malformed outputs—especially when the target format (e.g., JSON) is complex.  Second, we must maintain \textit{referential integrity}: references to the same entity instance must remain consistent \textit{across} related tables. For example, a tuple in the \texttt{Trip} table may refer to a destination (e.g., \textit{Rome}) and a traveler (e.g., \textit{Sophia}), who also appears in the \texttt{Traveler} table. Here, the traveler ID used in the \texttt{Trip} table must match the primary key of the corresponding tuple in the \texttt{Traveler} table (Fig. \ref{fig:schema}).
\\\noindent\textbf{Approach.} Before delving into the details, we remind readers that \systemname~has three possible inputs for table population: (1) text alone, (2) text with symbolic triplets, and (3) text with schema-aligned triplets. Including all three in a single prompt increases context length and can degrade output quality. Instead, each source is used independently as input to the prompt, and the resulting tuples are later \textit{combined}. This is akin to \textit{ensemble learning} in ML~\cite{polikar2012ensemble}, allowing us to leverage the complementary strengths of each input. 

We now describe the process of table population. To address the value-alignment challenge, we use a structured format that is \textit{incrementally generatable} by the LLM. Instead of emitting the entire structure at once, the format supports iterative generation, which reduces formatting errors. We ensure referential integrity by incorporating carefully chosen guidelines in the prompt that is compatible with the above format. 
In particular, we leverage  \textit{tool use} in LLMs \cite{Qu_2025} by introducing a lightweight tool \textbf{extract} that outputs one structured record at a time according to a given schema. This approach helps the LLM remain consistent with the expected output format.



\begin{figure}[H]
  \vspace{-1em}  \centering
    \includegraphics[width=1\linewidth]{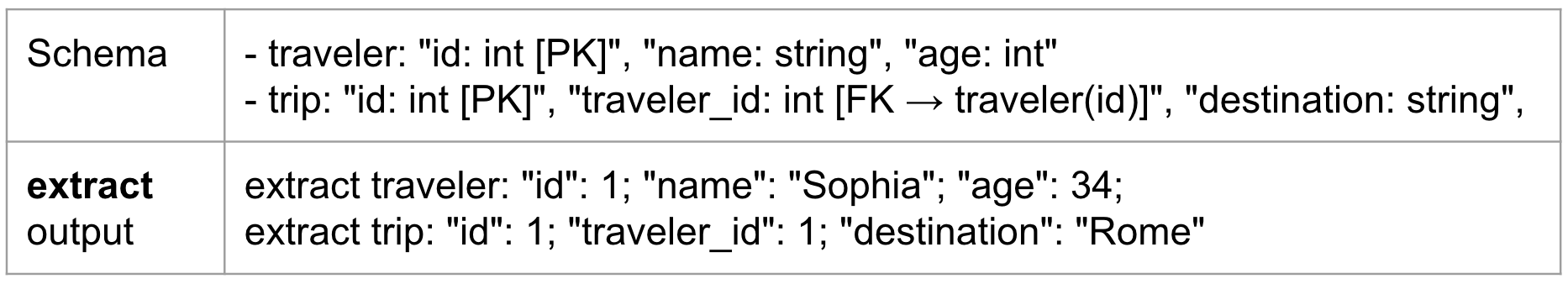}
    \label{fig:enter-label}
\end{figure}
\vspace{-2.5em}
After generating the records, we parse the output to extract each column-value pair for every tuple.


\subsection{Database Materialization}
\label{subsec-db-gen}
\textbf{Challenge.} 
A na\"ive approach is to prompt LLMs with all prior schema and value information to generate the corresponding SQL \texttt{INSERT} statements directly. However, this method is both inefficient and error-prone. We observe that this is akin to a ``program synthesis'' task---it not only requires the production of a large number of redundant tokens, which can be costly; but is also brittle to slight mistakes (e.g., a slightly-malformed SQL statement will produce execution errors). 
\\\noindent\textbf{Approach.} Instead, we observe that the required SQL statements are well-defined---creating specific tables and then inserting the corresponding tuples to these tables. Therefore, we decouple the materialization step from the LLM by parsing the model's output from the previous stage to \textit{programmatically} construct executable SQL code. Specifically, we generate \texttt{CREATE TABLE} and \texttt{INSERT INTO} statements (as shown in Fig. \ref{fig:e2e}) which are executed on a local SQLite instance to instantiate the database.  This separation enables deterministic parsing, ensuring syntactically correct SQL statements.

\section{Evaluation Setup}\vspace{-0.2cm}
\noindent\textbf{Dataset.} The \taskname~task requires a text document paired with a ground-truth relational database—however, no existing benchmarks directly support this. To fill this gap, we introduce an automated dataset creation pipeline: starting from relational databases or CSV files (using column names and tuple values as ground truth), we prompt an LLM to generate textual descriptions of the tuples, which serve as the input for \taskname. Using this approach, we construct two datasets: (1) \textbf{BIRD Dataset}—covering six domains from the BIRD Text2SQL benchmark~\cite{bird_benchmark}; and (2) \textbf{Kaggle Dataset}—containing CSV files from three domains (\textit{tourism, education, finance}) \cite{traveler_trip_2023, education_kaggle, fintech_ltv_2023}, which reflect more user-centric, realistic data often missing in BIRD. Table~\ref{tab:dataset} summarizes the dataset statistics. We categorize the text difficulty as \textbf{easy} (e.g., \textit{Tourism, Finance}), \textbf{medium} (e.g., \textit{Education, California Schools}), or \textbf{hard} (e.g., \textit{Mental Health, Superheroes}), based on domain complexity, record sparsity, and LLM-induced verbosity.
\begin{table}[ht]
    \centering
    \resizebox{\columnwidth}{!}{%
\begin{tabular}{lccccccccc}
\toprule
\textbf{Domain} & \multicolumn{3}{c}{\textbf{Kaggle (24 tables/domain)}} & \multicolumn{6}{c}{\textbf{BIRD (24 tables/domain)}} \\
\cmidrule(lr){2-4} \cmidrule(lr){5-10}
& Tourism & Education & Finance & Calif. Schools & Superhero & Books & Comp. Student & Mental Health & Authors \\
\midrule
Cols/Table   & 12  & 8   & 10  & 26  & 9  & 7  & 6  & 2  & 5 \\
Vals/Table   & 60  & 40  & 50  & 130 & 45 & 35 & 30 & 10 & 25 \\
\midrule
\multicolumn{9}{r}{\textbf{Overall Total Values}} & \textbf{10{,}200} \\
\bottomrule
\end{tabular}

    }\vspace{-0.2cm}
    \caption{Dataset statistics}\vspace{-0.5cm}
    \label{tab:dataset}
\end{table}
\\\noindent \textbf{Models.}
We test five state-of-the-art models: \textsc{GPT-4o} \cite{openai2024gpt4o}, \textsc{DeepSeek-v2.5} \cite{deepseek2024v25}, \textsc{Claude 3.7 Sonnet} \cite{anthropic2024claude3}, \textsc{LLaMA-3-8B-Instruct} \cite{meta2024llama3}, and \textsc{Qwen3-8B} \cite{alibaba2024qwen3b}.\\
\noindent\textbf{Metrics.} We propose a suite of novel metrics for a principled evaluation of the \taskname~task, which are summarized in Table \ref{tab:metrics-summary}.\\
\noindent\textit{Schema Evaluation.} We evaluate the quality of generated database schemas along three dimensions: \textbf{entity coverage}, \textbf{primary key coverage}, and \textbf{foreign key coverage}. Entity coverage assesses whether each column from the ground truth is represented in the generated schema. A column is considered covered if there exists a semantically equivalent column (based on cosine similarity between column names) in the output. 
Primary key coverage checks whether each generated table defines at least one primary key, while foreign key coverage evaluates whether all foreign keys correctly reference primary keys in valid, related tables within the schema. The last two metrics assess syntactic constraints that are essential for the correctness of relational database schemas.

\begin{table*}[t]
\footnotesize
\centering

{
\renewcommand{\arraystretch}{1.4}
\resizebox{0.9\textwidth}{!}{
\begin{tabular}{llll}
\toprule
& \textbf{Evaluation Metrics} & \bf{Definition} & \bf{Formula}                                                                                                                                             \\ \hline
\multirow{3}{*}{Schema} & Entity Coverage Score (\texttt{ECS})                                                         & Avg. max cosine similarity betn. GT \& DB columns & $ \frac{1}{N} \sum_{i=1}^N \max_{j \in [1, M]} \, \mathtt{cos\_sim}(c_i, \hat{c}_j) $ \\
& Primary Key Coverage (\texttt{PKC})                                                          & \% of tables with a defined primary key             & $ \nicefrac{\mathtt{\#Tables\;with\;PK}}{\mathtt{\#Tables}} $ \\
& Foreign Key Coverage (\texttt{FKC})                                                          & \% of tables whose foreign keys refer valid primary keys   & $ \nicefrac{\mathtt{\#Tables\;with\;valid\;FK}}{\mathtt{\#Tables}}
 $\\ \hline
\multirow{5}{*}{Tuple}& Database Construction Success Rate (\texttt{DBR}) & \% of successfully generated databases                & $\nicefrac{\mathtt{\#Generated\;DB}}{\mathtt{\#Text\;Documents}}
 $ \\
& Tuple Coverage (\texttt{TC})                                                                   & \% of GT tuples present in DB            & $\nicefrac{|\mathcal{R}_{\mathtt{GT}} \cap \mathcal{R}_{\mathtt{DB}}|}{|\mathcal{R}_{\mathtt{GT}}|}$

 \\
& Value Coverage (\texttt{VC})                                                                 & \% of GT values present in DB   & $\nicefrac{|\mathcal{V}_{\mathtt{GT}} \cap \mathcal{V}_{\mathtt{DB}}|}{|\mathcal{V}_{\mathtt{GT}}|}$
 \\
& Column Consistency (\texttt{CC})                                                             & \% of GT values present in DB in correct columns      & $\nicefrac{|\mathcal{V}_{\mathtt{GT}} \cap \mathcal{V}_{\mathtt{DB}}|}{ |\mathcal{V}_{\mathtt{GT}}|}$, where $\mathcal{V}_{\mathtt{GT}}, \mathcal{V}_{\mathtt{DB}} \in \mathtt{col}$ \\

& Ref. Integrity \% (\texttt{RRIR})                                                    & Avg. tuple completeness after FK joins (non-null ratio)       & $\mathtt{RRIR} = \frac{1}{N} \sum_{i=1}^N \frac{\mathtt{\# non-null\;values\;in\;tuple\;} i}{\mathtt{\# total\;values\;in\;tuple\;} i}$ \\
\bottomrule
\end{tabular}
}
}
\vspace{-1em}
\caption{Novel evaluation metrics for \taskname: GT denotes ground truth and DB denotes the generated databases. }
\label{tab:metrics-summary}
\vspace{-0.6cm}
\end{table*}
\noindent\textit{Tuple Evaluation.}
Relational databases store data across multiple tables; therefore, evaluating the quality of such databases requires a holistic view that goes beyond individual tables or isolated values. To enable a principled evaluation, we flatten the schema into a single table—commonly referred to as a denormalized table~\cite{db-book}—by performing a \texttt{JOIN} across all tables. In our databases, each table maintains a many-to-one or one-to-one relationship with a central table, enabling this complete \texttt{JOIN} of the entire schema. This consolidated table captures complete entity-relationship instances in a unified format. We generate two denormalized tables: one from the ground-truth database and one from the database produced by \systemname. The two are then compared to assess the accuracy of the generated database.

We propose five novel metrics to evaluate the quality of the generated tuples along two dimensions: syntactic and semantic validity. Syntactic validity assesses whether the generated databases adhere to correct structural and relational rules. It is measured using: (1) \textbf{Database Construction Success Rate}, which measures the percentage of generated SQL statements that successfully materialize into databases with at least one non-null tuple, (2) \textbf{Referential Integrity Rate} \texttt(RRIR), which measures the fraction of foreign-key joins that yielded valid (non-null) tuples. 

Semantic validity evaluates the comprehensiveness and correctness of the values populated. It is measured using: (1) \textbf{Tuple Coverage}, which measures the fraction of the ground truth tuples recovered; (2) \textbf{Value Coverage}, which measures the fraction of ground truth values populated; and (3) \textbf{Column Consistency}, which checks whether each value appears in its correct column.\\
\noindent\textbf{Baseline.} Our \taskname~task is novel, and prior work targets fundamentally different objectives (see Sec.\ref{sec:task}), making direct comparison infeasible. To address this, we design a tailored baseline: using zero-shot prompting, we generate \texttt{CREATE} \texttt{TABLE} and \texttt{INSERT}  \texttt{INTO} SQL statements directly from the input text, then execute them in SQLite to instantiate the database. Prompt details are in Appendix~\ref{appendix:prompts}.
\vspace{-6mm}
\section{Experiments and Analysis}\vspace{-0.2cm}
We evaluate the performance of \systemname~based on the following three research questions (RQs):
\begin{packeditemize} \vspace{-0.2cm}
\item \textbf{RQ1.} Can \systemname~generate a high-quality relational schema?
\item \textbf{RQ2.} Can \systemname~generate 
accurate relational tuples to populate the tables? 
\item \textbf{RQ3.} How do \systemname's design choices affect performance?
\end{packeditemize}\vspace{-0.5cm}
\begin{table*}[!]
\scriptsize
\centering
\setlength{\tabcolsep}{2pt}

\small
\scalebox{0.8}{\begin{tabular}{llcccccccccccc}
\toprule
\textbf{Model} & \textbf{Prompt} & \multicolumn{3}{c}{\textbf{Easy}} & \multicolumn{3}{c}{\textbf{Medium}} & \multicolumn{3}{c}{\textbf{Hard}} & \multicolumn{3}{c}{\textbf{Avg.}} \\
 & & \scriptsize{\texttt{ECS}(\%)} & \scriptsize{\texttt{PKC}(\%)} & \scriptsize{\texttt{FKC}(\%)} & \scriptsize{\texttt{ECS}(\%)} & \scriptsize{\texttt{PKC}(\%)} & \scriptsize{\texttt{FKC}(\%)} & \scriptsize{\texttt{ECS}(\%)} & \scriptsize{\texttt{PKC}(\%)} & \scriptsize{\texttt{FKC}}(\%) & \scriptsize{\texttt{ECS}}(\%) & \scriptsize{\texttt{PKC}(\%)} & \scriptsize{\texttt{FKC}(\%)} \\
\midrule
\multirow{1}{*}{\textsc{Claude 3.7 Sonnet}} 
    & Direct & 86.2 & 100 & 100 & 80.7 & 100 & 100 & 45.3 & 100 & 100 & 70.7 & 100 & 100 \\
\midrule
\multirow{2}{*}{\textsc{LlaMa-8B Instruct}} 
    & Direct & 80.4 & 100 & 100 & 78.6 & 100 & 100 & 55.4 & 100 & 100 & 71.5 & 100 & 100 \\
    & CoT    & 95.8 & 100 & 100 & 76.5 & 100 & 100 & 62.7 & 100 & 100 & 78.4 & 100 & 100 \\
\midrule
\multirow{2}{*}{\textsc{Deepseek v2.5}} 
    & Direct & -- & -- & -- & 28.3 & 33.33 & 33.3 & 34.5 & 50 & 50 & 20.9 & 27.8 & 27.8 \\
    & CoT    & 86.9 & 100 & 100 & 84.2 & 100 & 100 & 65.1 & 100 & 100 & 78.8 & 100 & 100 \\
\midrule
\multirow{2}{*}{\textsc{GPT-4o}} 
    & Direct & 90.5 & 100 & 100 & 79.4 & 94.4 & 100 & 62.6 & 100 & 100 & 77.5 & 98.2 & 100 \\
    & CoT    & 93.0 & 100 & 100 & 80.0 & 100 & 66.7 & 63.2 & 100 & 100 & 78.7 & 100 & 88.9 \\
\bottomrule
\end{tabular}}

\vspace{-0.1cm}\caption{Schema evaluation:  Entity (\texttt{ECS}), Primary Key (\texttt{PKC}) and Foreign Key (\texttt{FKC}) coverage scores. ``–'': schema generation failures that violate the requested structure in our prompts.  \textsc{Claude}-CoT and \textsc{Qwen-8B} are omitted due to such failures. }
\label{tab:schema-coverage}\vspace{-0.4cm}
\end{table*}


\subsection{RQ1: Schema Evaluation} \vspace{-0.1cm}
As described in Sec.\ref{subsec-schema-gen}, we evaluate two prompting strategies for schema generation: Direct and Chain-of-Thought (CoT). Table~\ref{tab:schema-coverage} summarizes the results. We only consider schemas that match the format specified in the prompt, as this is required for \systemname~to process them later. We evaluate both syntactic validity—using primary key coverage (\texttt{PKC}) and foreign key coverage (\texttt{FKC})—and semantic validity, using entity coverage (\texttt{ECS}). We first highlight general observations across all three metrics, followed by specific analysis. 
Overall, CoT consistently outperforms Direct across difficulty levels; except \textsc{Claude}, which performs better with Direct but struggles with CoT due to format violations, likely due to overthinking \cite{cot-is-bad}. \textsc{Qwen-8B} consistently fails to produce valid schemas, likely due to poor support for structured output tasks \cite{qwen-is-bad}.  
\\\noindent\textbf{Syntactic Validity.}
We observe that most CoT-based generations achieve full \texttt{PKC} and \texttt{FKC}, except \textsc{GPT}, which drops to 66.67\% \texttt{FKC} in the medium dataset. This is because \textsc{GPT} occasionally generates a single table with no foreign key, when the text contains only a few entities.
\\\noindent\textbf{Semantic Validity.} For entity coverage \texttt{ECS}, \textsc{DeepSeek} with CoT performs the best, followed by \textsc{LLaMa-8B} and \textsc{GPT}--which show minor drops due to their tendency to generate paraphrased column names (e.g., “heritage” or “ethnicity” instead of “race”), whereas \textsc{DeepSeek} aligns more closely with the ground truth. In terms of performance across domains (Appendix~\ref{appendix:experiments}), \textsc{DeepSeek} achieves the highest entity coverage in the \textit{Education} domain (91.08\%) and the lowest in the \textit{Mental Health} domain (38.97\%). The ground truth of the latter has complex column names, such as ``questiontext" and ``answertext", suggesting that domain complexity significantly affects the quality of the generated schema. 

\vspace{-2.3mm}

\subsection{\textbf{RQ2:} Tuple Evaluation} 


\noindent\textbf{Syntactic Validity.}
Table~\ref{tab:pctg_correct_db} reports the Database Construction Success Rate (\texttt{DBR}) and the improvement in Referential Integrity Rate (\texttt{RRIR}) over the baseline. We highlight three observations. First, \systemname~achieves perfect \texttt{DBR} (100\%) across all models and difficulty levels, except for using \textsc{DeepSeek} on hard examples, where it drops slightly to 98\%. This indicates the robustness of \systemname~in consistently generating syntactically valid databases. In contrast, the baseline \texttt{DBR} varies widely—from as low as 9.7\% (\textsc{GPT}) to 58.2\% (\textsc{Claude}) on average.
Next, we turn to referential integrity. We note that \systemname’s \texttt{RRIR} is a conservative (lower-bound) estimate, since records with missing values in the ground truth are treated as invalid under our metric. 
Nevertheless, \systemname~still achieves significant improvements over the baseline. For example, \textsc{GPT} exhibits the highest improvement ($46.59\times$ on easy examples). \textsc{Qwen-8B} also achieve notable average improvements of $3.52\times$. Although \textsc{LLaMA-8B} achieves perfect \texttt{DBR}, its \texttt{RRIR} does not improve on the medium dataset, suggesting its baseline already exhibits relatively strong referential integrity.
\begin{table}[tbh]
    \centering
    \footnotesize
    \setlength{\tabcolsep}{2pt} 





\scalebox{0.8}{\begin{tabular}{llccc}
\toprule
& & \multicolumn{2}{c}{\textbf{DBR(\%)}} & \textbf{RRIR} \\
& & \textbf{\systemname} & \textbf{Baseline} & \scriptsize{\textbf{Improvement Factor}}\\
\midrule

\multirow{4}{*}{\makecell[l]{\textsc{Claude 3.7}\\\textsc{Sonnet}}}
& Easy    & 100.0 & 63.2 & 1.56$\times$ \\
& Medium  & 100.0 & 63.4 & 1.10$\times$ \\
& Hard    & 100.0 & 48.1 & 1.41$\times$ \\ 
\cmidrule(lr){2-5}
& Average & 100.0 & 58.2 & 1.40$\times$ \\
\midrule

\multirow{4}{*}{\makecell[l]{\textsc{Deepseek}\\\textsc{v2.5}}}
& Easy    & 100.0 & 23.2 & 4.44$\times$ \\
& Medium  & 100.0 & 42.4 & 1.70$\times$ \\
& Hard    & 98.0 & 40.3 & 1.87$\times$ \\
\cmidrule(lr){2-5}
& Average & 99.3 & 35.3 & 1.80$\times$ \\
\midrule

\multirow{4}{*}{\makecell[l]{\textsc{GPT-4o}\\}}
& Easy    & 100.0 & 2.0 & 46.59$\times$ \\
& Medium  & 100.0 & 6.1 & 12.09$\times$ \\
& Hard    & 100.0 & 21.0 & 2.63$\times$ \\
\cmidrule(lr){2-5}
& Average & 100.0 & 9.7 & 13.93$\times$ \\
\midrule

\multirow{4}{*}{\makecell[l]{\textsc{Qwen3}\\\textsc{-8B}}} 
& Easy    & 100.0 & 23.5 & 4.42$\times$ \\
& Medium  & 100.0 & 32.2 & 2.52$\times$ \\
& Hard    & 100.0 & 10.4 & 6.83$\times$ \\
\cmidrule(lr){2-5}
& Average & 100.0 & 22.0 & 3.52$\times$ \\
\midrule

\multirow{4}{*}{\makecell[l]{\textsc{LLaMA-3}\\\textsc{8B-Instruct}}}
& Easy    & 100.0 & 63.2 & 1.54$\times$ \\
& Medium  & 100.0 & 64.5 & 1.00$\times$ \\
& Hard    & 100.0 & 40.1 & 1.87$\times$ \\
\cmidrule(lr){2-5}
& Average & 100.0 & 55.9 & 1.64$\times$ \\
\bottomrule
\end{tabular}}

\vspace{-0.4cm}    \caption{Database Construction Success Rate (\%) and the improvement factor in Referential Integrity Rate in \systemname compared to the baseline.} 
    \label{tab:pctg_correct_db} \vspace{-0.3cm}  
\end{table}
\\\noindent\textbf{Semantic Validity.}
Table~\ref{tab:db-coverage} reports Tuple Coverage (\texttt{TC}), Value Coverage (\texttt{VC}), and Column Consistency (\texttt{CC}) with three findings. First, \systemname consistently outperforms the baseline across all models and metrics. Notably, all 8B-parameter models (\textsc{LLaMa-8B, Qwen-8B}) under \systemname significantly outperform all larger model baselines (\textsc{GPT, Claude, DeepSeek}). In particular, although \textsc{Qwen-8B}'s baseline lags behind those of \textsc{Claude} and \textsc{DeepSeek}, its performance under \systemname surpasses them—highlighting the effectiveness of our approach. Second, on average, all models using \systemname achieve high \texttt{TC} ($\geq$0.95) and strong \texttt{VC/CC} ($\geq$0.70), with \textsc{GPT} showing the largest improvement over its baseline (17.75$\times$ improvement on \texttt{CC}). This is primarily because failed database generations are assigned zero scores, and as shown in Table~\ref{tab:pctg_correct_db}, \textsc{GPT} performs poorly in database construction under the baseline setting. Third, even for models with relatively strong baseline performance, such as \textsc{LLaMa-8B}, \systemname improves \texttt{VC} and \texttt{CC} by $4.1\times$ and $5.5\times$ on hard examples, respectively. 

\vspace{-0.3cm}
\begin{table}[ht]
\centering
\scriptsize 
\setlength{\tabcolsep}{2pt} 

\scalebox{0.9}{\begin{tabular}{llcccccc}
\toprule
\multirow{2}{*}{\rotatebox{90}{\textbf{Model}}} & \multirow{2}{*}{\rotatebox{90}{\textbf{Diff.}}} & \multicolumn{3}{c}{\systemname} & \multicolumn{3}{c}{Baseline} \\
\cmidrule(lr){3-5} \cmidrule(lr){6-8}
 &  & \texttt{TC}(\%) & \texttt{VC}(\%) & \texttt{CC}(\%) & \texttt{TC}(\%) & \texttt{VC}(\%) & \texttt{CC}(\%) \\
\midrule

\multirow{4}{*}{\rotatebox{90}{\parbox{1.2cm}{\tiny{\textsc{Claude 3.7}}}}}
& Easy    & 100.0 (2.56$\times$) & 98.0 (4.67$\times$) & 98.0 (4.67$\times$) & 39.0 & 21.0 & 21.0 \\
& Med     & 98.0 (2.72$\times$) & 78.0 (6.00$\times$) & 74.0 (5.69$\times$) & 36.0 & 13.0 & 13.0 \\
& Hard    & 100.0 (2.44$\times$) & 63.0 (2.74$\times$) & 41.0 (3.73$\times$) & 41.0 & 23.0 & 11.0 \\
\cmidrule(lr){2-8}
& Avg     & 99.0 (2.61$\times$) & 80.0 (4.21$\times$) & 71.0 (4.73$\times$) & \cellcolor{gray!30}38.0 & \cellcolor{gray!30}19.0 & \cellcolor{gray!30}15.0 \\
\midrule

\multirow{4}{*}{\rotatebox{90}{\parbox{1.3cm}{\scalebox{0.9}{\tiny{\textsc{Deepseek-v2.5}}}}}}
& Easy    & 100.0 (5.88$\times$) & 96.0 (6.86$\times$) & 96.0 (6.86$\times$) & 17.0 & 14.0 & 14.0 \\
& Med     & 99.0 (3.54$\times$) & 80.0 (5.33$\times$) & 77.0 (5.50$\times$) & 28.0 & 15.0 & 14.0 \\
& Hard    & 95.0 (2.64$\times$) & 59.0 (2.57$\times$) & 39.0 (3.90$\times$) & 36.0 & 23.0 & 10.0 \\
\cmidrule(lr){2-8}
& Avg     & 98.0 (3.63$\times$) & 79.0 (4.65$\times$) & 71.0 (5.92$\times$) & \cellcolor{gray!30}27.0 & \cellcolor{gray!30}17.0 & \cellcolor{gray!30}12.0 \\
\midrule

\multirow{4}{*}{\rotatebox{90}{\parbox{0.9cm}{\tiny{\textsc{GPT-4o}}}}}
& Easy    & 100.0 (50.00$\times$) & 97.0 (48.50$\times$) & 97.0 (48.50$\times$) & 2.0 & 2.0 & 2.0 \\
& Med     & 99.0 (16.50$\times$) & 81.0 (16.20$\times$) & 77.0 (19.25$\times$) & 6.0 & 5.0 & 4.0 \\
& Hard    & 97.0 (6.47$\times$) & 61.0 (5.55$\times$) & 40.0 (6.67$\times$) & 15.0 & 11.0 & 6.0 \\
\cmidrule(lr){2-8}
& Avg     & 99.0 \textbf{(14.14$\times$)} & 80.0 \textbf{(13.33$\times$)} & 71.0 \textbf{(17.75$\times$)} & \cellcolor{gray!30}7.0 & \cellcolor{gray!30}6.0 & \cellcolor{gray!30}4.0 \\
\midrule

\multirow{4}{*}{\rotatebox{90}{\parbox{1.2cm}{\scalebox{0.8}{\tiny{\textsc{LlaMa3-8B-In.}}}}}}
& Easy    & 100.0 (1.82$\times$) & 95.0 (3.06$\times$) & 95.0 (3.17$\times$) & 55.0 & 31.0 & 30.0 \\
& Med     & 99.0 (1.83$\times$) & 79.0 (2.82$\times$) & 75.0 (3.00$\times$) & 54.0 & 28.0 & 25.0 \\
& Hard    & 100.0 (3.45$\times$) & 70.0 (4.12$\times$) & 44.0 (5.50$\times$) & 29.0 & 17.0 & 8.0 \\
\cmidrule(lr){2-8}
& Avg     & \cellcolor{gray!30}\textbf{100.0} (2.17$\times$) & \cellcolor{gray!30}\textbf{81.0} (3.24$\times$) & \cellcolor{gray!30}71.0 (3.38$\times$) & 46.0 & 25.0 & 21.0 \\
\midrule

\multirow{4}{*}{\rotatebox{90}{\parbox{1.2cm}{\tiny{\textsc{Qwen3-8B}}}}}
& Easy    & 100.0 (4.55$\times$) & 96.0 (5.05$\times$) & 96.0 (5.33$\times$) & 22.0 & 19.0 & 18.0 \\
& Med     & 98.0 (3.27$\times$) & 79.0 (3.16$\times$) & 79.0 (3.43$\times$) & 30.0 & 25.0 & 23.0 \\
& Hard    & 99.0 (14.14$\times$) & 51.0 (10.20$\times$) & 51.0 (17.00$\times$) & 7.0 & 5.0 & 3.0 \\
\cmidrule(lr){2-8}
& Avg     & \cellcolor{gray!30}99.0 (4.95$\times$) & \cellcolor{gray!30}76.0 (4.75$\times$) & \cellcolor{gray!30}\textbf{75.0} (5.00$\times$) & 20.0 & 16.0 & 15.0 \\
\bottomrule
\end{tabular}}

\vspace{-0.2cm}  \caption{Tuple evaluation via Tuple Coverage (\texttt{TC}), Value Coverage (\texttt{VC}) and Column Consistency (\texttt{CC}). Best scores and improvement factors across models in \textbf{bold}. \textcolor{gray}{Gray} indicates that \systemname on all 8B models outperforms  larger models. }
\vspace{-0.5cm}  
\label{tab:db-coverage}
\end{table}

\begin{table}[ht]
\centering
\scriptsize 
\setlength{\tabcolsep}{3pt} 

\scriptsize
\begin{tabular}{p{6mm}rcccccc}
\toprule
\tiny{\textbf{Model}}                              & \textbf{Diff.}   & \mOne (\%)   & \mTwo (\%)  & \mThree (\%) & \eOne (\%) & \eTwo (\%) & \eThree (\%) \\ 
& & \tiny{(1)}   & \tiny{(2)}  & \tiny{(3)} & \tiny{(1)+(2)} & \tiny{(1)+(3)} & \tiny{(1)+(2)+(3)} \\ 

\midrule
\multirow{4}{*}{\rotatebox{90}{\parbox{1.2cm}{\tiny{\textsc{Claude 3.7}}}}} 
& Easy    & 97.4 & 97.1 & 93.8 & 98.3 & 97.7 & 98.4 \\
& Med  & 68.2 & 74.6 & 74.1 & 77.3 & 77.5 & 78.2 \\
& Hard    & 51.7 & 58.4 & 51.3 & 60.7 & 60.5 & 63.1 \\
\cmidrule(lr){2-8}
& Avg & 72.4 & 76.7 & 73.1 & 78.8 & 78.6 & 79.9 \\ \midrule

\multirow{4}{*}{\rotatebox{90}{\parbox{1.2cm}{\tiny{\textsc{Deepseek-v2.5}}}}} 
& Easy    & 92.3 & 94.5 & 92.7 & 96.8 & 95.4 & 96.8 \\
& Med  & 76.7 & 68.1 & 69.3 & 79.6 & 80.2 & 80.4 \\
& Hard    & 54.8 & 42.9 & 35.7 & 57.4 & 57.1 & 59.3 \\
\cmidrule(lr){2-8}
& Avg & 74.6 & 68.5 & 65.9 & 77.9 & 77.6 & 78.8 \\ \midrule

\multirow{4}{*}{\rotatebox{90}{\tiny{\parbox{0.8cm}{\textsc{GPT-4o}}}}} 
& Easy    & 90.8 & 93.2 & 90.4 & 95.1 & 96.3 & 97.4 \\
& Med  & 75.3 & 69.7 & 68.6 & 80.4 & 81.2 & 81.3 \\
& Hard    & 50.6 & 41.8 & 51.7 & 56.3 & 59.4 & 61.6 \\
\cmidrule(lr){2-8}
& Avg & 72.2 & 68.2 & 70.2 & 77.3 & 79.0 & 80.1 \\ \midrule

\multirow{4}{*}{\rotatebox{90}{\parbox{1.3cm}{\scalebox{.7}{\textsc{LlaMa3-8B-In.}}}}} 
& Easy    & 89.3 & 74.8 & 61.5 & 95.2 & 94.4 & 95.5 \\
& Med  & 70.1 & 52.7 & 61.3 & 75.5 & 76.1 & 79.4 \\
& Hard    & 60.6 & 37.9 & 40.4 & 64.3 & 68.5 & 70.7 \\
\cmidrule(lr){2-8}
& Avg & 73.3 & 55.1 & 54.4 & 78.3 & 79.7 & 81.9 \\ \midrule

\multirow{4}{*}{\rotatebox{90}{\parbox{1.2cm}{\tiny{\textsc{Qwen3-8B}}}}} 
& Easy    & 92.1 & 92.6 & 72.5 & 96.4 & 96.1 & 96.4 \\
& Med  & 71.4 & 71.3 & 67.4 & 74.7 & 78.4 & 79.5 \\
& Hard    & 29.8 & 23.5 & 35.9 & 33.3 & 48.2 & 51.2 \\
\cmidrule(lr){2-8}
& Avg & 64.4 & 62.5 & 58.6 & 68.1 & 74.2 & 75.7 \\

\bottomrule
\end{tabular}

\vspace{-0.2cm}\caption{Impact of different value source. The first three columns represent individual prompt settings, while the last three correspond to post-generation ensembling. \eOne combines tuples generated from \mOne and \mTwo while \eTwo combines \mOne and \mThree. \systemname~combines outputs from all three prompts. }
\label{tab:db-coverage-all} \vspace{-0.5cm}
\end{table}


\subsection{RQ3: Impact of \systemname's Design Choices}\vspace{-0.1cm}
\label{sec:eval-design}

We now evaluate the impact of \systemname's design choices on value identification and table population. Recall that we consider three different prompts for table population based on their input source:
(1) text only (\mOne),
(2) text with symbolic triplets (\mTwo), and
(3) text with schema-aligned triplets (\mThree).
\systemname~combines the rows generated from all three prompts. 
Table~\ref{tab:db-coverage-all} evaluates how these different value sources affect the quality of the generated tuples, with the following observations. 

First, using triplets significantly improves value coverage compared to extracting them from the text alone. This is evident from the observation that  \systemname outperforms \mOne by 5–12\%. 

Second, we examine how to best incorporate the triplets: whether to concatenate them with the input text in a \textit{single} prompt,  or to generate tuples separately and combine them post-hoc (ensembling). \systemname~adopts the latter strategy, and our results support this choice. Specifically, in the individual prompt setting, \mOne outperforms both \mTwo and \mThree in all but one case (\textsc{CLAUDE}). In contrast, the ensemble approaches (\eOne, \eTwo and \systemname) consistently outperform all the individual prompts. This suggests that including triplets directly in the input prompt increases context length, which degrades model performance—likely due to context window saturation \cite{llm-is-bad1}.

Finally, we evaluate our design choice of combining triples generated from symbolic tools and schema-aligned triplets from LLMs. Overall, \eTwo outperforms \eOne across most models on average, except for \textsc{Claude} and \textsc{Deepseek}. \systemname consistently yields the best score, indicating that each source captures \textit{complementary} information. LLM-generated triplets are schema-aware and can correctly group multi-word values under the correct columns (e.g., mapping “car rental” to the \textit{transportation mode} column, whereas symbolic tools only captured “car”). However, LLMs sometimes paraphrase values (e.g., “low income” to “modest income”), whereas symbolic tools extract values verbatim, yielding closer alignment to the input. 
\vspace{-0.1cm}
\section{Related Work}\vspace{-0.1cm}
\textbf{Summarizing Structures.} \textit{Text-to-table} generation \cite{wu2022texttotable, sundar2024gtbls, li2023seq2seqset, deng2024texttupletableinformationintegrationtexttotable, Aroravldb, jain2024structsumgenerationfastertext} projects explore sequence-to-sequence modeling, LLM prompt engineering, and structured summarization techniques. However, they can only generate flat tables, and cannot capture the relational database model in our work. 
\\\noindent\textbf{Manipulating Existing Databases.} The goal of these projects is to leverage LLMs to interact with existing relational databases—such as to generate SQL queries from text \cite{hong2024next,pang2020ratsql}, or to update them using natural language \cite{jiao-etal-2024-text2db}. However, none of these works can synthesize a relational database from scratch, which is what \systemname tackles. 
\\\noindent\textbf{Non-LLM Approaches.} Prior to LLMs, integrating text into relational structures relied on traditional pipelines that combine information extraction, schema induction, and entity linking \cite{zhang2016deepdive,smith2022lillie,zhang2019openki}. These methods rely on  statistical or symbolic techniques, but required domain-specific heuristics and did not generalize to noisy or diverse input text. 
\vspace{-0.2cm}\section{Conclusion}\vspace{-0.2cm}
In this work, we have introduced a novel task of synthesizing relational databases from text, called \taskname. We have also developed a framework, \systemname, designed to solve \taskname~tasks. \systemname has a neurosymbolic pipeline, with each stage incorporating specialized techniques for the task. Our experiments show \systemname significantly outperforms baseline solutions across diverse datasets. 
\subsection*{Limitations}
While we provide extensive evaluation of \systemname on our benchmark, we leave comparisons with few-shot baselines and fine-tuned models for future work. Additionally, our current evaluation method is limited to user-centric text documents—that is, datasets where the generated database features a single central table to which all other tables relate, enabling a comprehensive SQL \texttt{JOIN}. This approach may not generalize to more complex schemas lacking such a central entity.  Also, our evaluation relies primarily on public datasets, which may not fully capture the complexities of open-ended, real-world text; future work should extend evaluation to diverse, naturally occurring data sources.



\subsection*{Ethics Statement}
All datasets used in this work are publicly available and released under open licenses. The tools and models employed are authorized for research purposes and have been used in accordance with their intended terms. Detailed license information is provided in Appendix~\ref{appendix:artifact}. All experiments were performed strictly for research and evaluation.

Because our study requires user-centric documents for schema generation and value mapping evaluation, anonymization was not feasible without significantly compromising data integrity. To the best of the authors’ knowledge, this research does not introduce any ethical risks beyond those already associated with the original datasets.

Since \systemname uses large language models (LLMs) to synthesize databases, and LLMs are known to occasionally produce hallucinated or inaccurate content, there are potential risks when applying \systemname in sensitive domains without human oversight. Careful review and verification are recommended before deploying the system in high-stakes or privacy-critical applications.

\bibliography{anthology,custom}

\clearpage
\appendix

\begin{appendix}

\section{Definitions}
\label{appendix:definitions}
\begin{enumerate}
    \item \textbf{Canonical Join Query:} The canonical join of the database schema is the natural join of all the relations in the schema. \cite{db-book}
    \item \textbf{Primary Key:} A primary key is a set of one or more attributes that uniquely identifies a tuple within a relation. No attribute in the primary key can have a null value. \cite[Section 3.3.2]{silberschatz2020}
    \item \textbf{Foreign Key:} A foreign key is an attribute, or a set of attributes, in one relation that references the primary key of another relation. It ensures referential integrity between the two relations. \cite[Section 3.4]{silberschatz2020}
    \item \textbf{Referential Integrity:} Referential integrity is a property of a relational database that ensures that every foreign key value in a child table either matches a valid primary key in the referenced parent table or is null (if allowed). It guarantees that relationships between tables remain consistent. \cite[Section 3.4]{silberschatz2020}
\end{enumerate}

\section{Dataset}
\label{appendix:dataset}
\begin{figure*}
    \centering
    \includegraphics[width=1\linewidth]{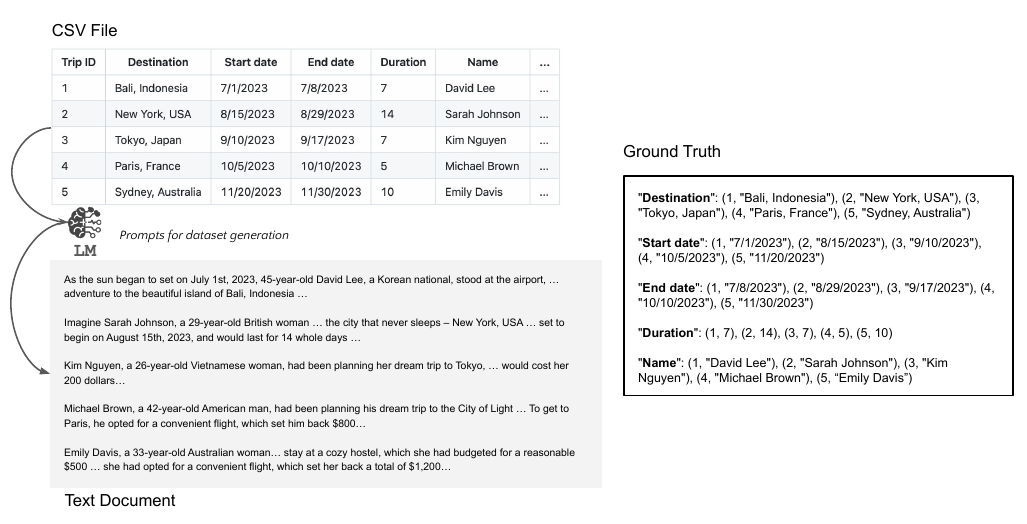}
    \caption{Our dataset generation process}
    \label{fig:dataset}
\end{figure*}

Our dataset generation approach is illustrated in Fig. \ref{fig:dataset}. For the BIRD dataset, we flatten each multi-table database obtained from the BIRD benchmark \cite{bird_benchmark} into a single table by joining related tables. Then, using the \textsc{LLaMA-8B-Instruct} model (see prompts in Appendix~\ref{appendix:prompt-dataset}), we generate a natural language sentence for each row. Five consecutive sentences are concatenated to create a paragraph-style input document. The same approach is applied to the Kaggle datasets \cite{traveler_trip_2023, fintech_ltv_2023, education_kaggle}.

\section{Metrics}
\label{appendix:metrics}
We assess the quality of both the generated schema and its instantiated content using a suite of novel evaluation metrics that capture structural correctness, semantic alignment, and data fidelity, providing a comprehensive measure of generation quality.

\subsubsection{Schema Evaluation}

We evaluate the quality of generated database schemas using three complementary metrics: \texttt{Entity Coverage Score (ECS)} for column-level semantic alignment, \texttt{Primary Key Coverage (PKC)} for schema completeness, and \texttt{Foreign Key Coverage (FKC)} for referential integrity.

\noindent
\textbf{\texttt{Entity Coverage Score (ECS)}} evaluates how well the predicted schema recovers the ground truth column names. Let \( \{ c_1, \ldots, c_N \} \) be the ground truth column names and \( \{ \hat{c}_1, \ldots, \hat{c}_M \} \) be the predicted columns. For each ground truth column \( c_i \), we compute its cosine similarity with every predicted column \( \hat{c}_j \) and select the highest similarity. \texttt{ECS} is the average of these maximum scores:
\begin{equation}
    \mathtt{ECS} = \frac{1}{N} \sum_{i=1}^N \max_{j \in [1, M]} \, \mathtt{cos\_sim}(c_i, \hat{c}_j)
\end{equation}
where cosine similarity is computed as:
\begin{equation}
    \mathtt{cos\_sim}(u, v) = \frac{u \cdot v}{\|u\| \|v\|}
\end{equation}
This metric captures the best semantic match for each ground truth column using SentenceTransformer embeddings (\texttt{all-MiniLM-L6-v2}).

\noindent
\textbf{\texttt{Primary Key Coverage (PKC)}} measures how well the generated schema supports tuple-level uniqueness by checking whether primary keys are defined. 
\texttt{PKC} is defined as:
\begin{equation}
    \mathtt{PKC} = \frac{{\mathtt{Num\_PK}}}{\mathtt{Num\_tables}}
\end{equation}
Here, \( \mathtt{Num\_PK} \) is the number of generated tables that define at least one primary key, and \( \mathtt{Num\_tables} \) is the total number of generated tables.
This metric reflects the model’s ability to generate structurally valid tables that enforce row-level uniqueness through primary keys.

\noindent
\textbf{\texttt{Foreign Key Coverage (FKC)}} assesses the extent to which the generated schema maintains referential integrity across tables. 
\texttt{FKC} is defined as:
\begin{equation}
    \mathtt{FKC} = \frac{ \mathtt{Num\_FK}_{\mathtt{Valid}} }{ \mathtt{Num\_FK} }
\end{equation}
Here, \( \mathtt{Num\_FK}_{\mathtt{Valid}} \) is the number of foreign keys that correctly reference existing primary keys, and \( \mathtt{Num\_FK} \) is the total number of generated foreign keys.
This metric evaluates the model’s ability to establish valid inter-table relationships, ensuring that foreign keys point to legitimate primary key targets.





\subsubsection{Database Evaluation}

We use five evaluation metrics to assess how well the generated database reconstructs the ground truth data: \texttt{Database Construction Success Rate (DBR)}, \texttt{Referential Integrity Rate (RRIR)}, \texttt{Tuple Coverage (TC)}, \texttt{Value Coverage (VC)}, and \texttt{Column Consistency (CC)}. 

\noindent
\texttt{Database Construction Success Rate (DBR)} captures the percentage of successfully generated databases from text documents.
\begin{equation}
    \mathtt{DBR} = \frac{\mathtt{\#Generated\;DB}}{\mathtt{\#Text\;Documents}} 
\end{equation}

\noindent
\textbf{\texttt{Referential Integrity Rate (RRIR)}} captures whether foreign key joins result in meaningful, non-sparse rows during execution. Let \( \mathcal{D} \) be the set of evaluated databases, and let each database \( d \in \mathcal{D} \) produce a set of rows \( \mathcal{R}_d \) from a canonical foreign key join. For each row \( r \in \mathcal{R}_d \), let \( n_{\mathtt{total}}(r) \) be the number of columns, and \( n_{\mathtt{null}}(r) \) the number of columns with null values. The per-database score is:
\begin{equation}
    \mathtt{RRIR}(d) = \frac{1}{|\mathcal{R}_d|} \sum_{r \in \mathcal{R}_d} \left(1 - \frac{n_{\mathtt{null}}(r)}{n_{\mathtt{total}}(r)}\right)
\end{equation}
The overall score across all databases is:
\begin{equation}
    \mathtt{RRIR} = \frac{1}{|\mathcal{D}|} \sum_{d \in \mathcal{D}} \mathtt{RRIP}(d)
\end{equation}
This metric provides a practical signal of referential soundness during execution by quantifying the completeness of joined rows in terms of non-null content.

\noindent
\textbf{\texttt{Tuple Coverage (TC)}} quantifies how many ground truth rows are recovered through canonical joins. Let $\mathcal{R}_{\mathtt{GT}}$ be the set of primary keys from the ground truth database, and $\mathcal{R}_{\mathtt{join}}$ be the set of primary keys resulting from the canonical join query over the generated database. Then:
\begin{equation}
    \mathtt{TC} = \frac{|\mathcal{R}_{\mathtt{GT}} \cap \mathcal{R}_{\mathtt{join}}|}{|\mathcal{R}_{\mathtt{GT}}|}
\end{equation}
This metric reflects the row-level reconstruction accuracy.

\noindent
\textbf{\texttt{Value Coverage (VC)}} measures the proportion of ground truth cell values that are accurately recovered in the predicted database. A predicted value \( \hat{v} \) is considered a match to a ground truth value \( v \) if:
\begin{itemize}
    \item For numeric values: \( |v - \hat{v}| < 10^{-2} \) (i.e., absolute difference less than 0.01).
    \item For textual values: the cosine similarity between embeddings satisfies \( \mathtt{cos\_sim}(v, \hat{v}) > 0.8 \).
\end{itemize}

Let \( \mathcal{V}_{\mathtt{GT}} \) be the set of all ground truth values, and \( \mathcal{V}_{\mathtt{DB}} \) be the set of predicted values matched to ground truth under the criteria above. Then VC is defined as:

\begin{equation}
    \mathrm{VC} = \frac{|\mathcal{V}_{\mathtt{GT}} \cap \mathcal{V}_{\mathtt{DB}}|}{|\mathcal{V}_{\mathtt{GT}}|}
\end{equation}

This ratio reflects the overall proportion of correctly reconstructed cell values, incorporating both numeric precision and semantic similarity for text.

\vspace{1em}

\noindent
\textbf{Column Consistency (CC)} quantifies the proportion of matched values that appear under the correct column names in the predicted database. A column name in the prediction is considered correct if its semantic similarity with the corresponding ground truth column name exceeds a threshold of 0.7, i.e.,

\[
\mathtt{cos\_sim}(\text{col}_{\mathtt{GT}}, \text{col}_{\mathtt{DB}}) > 0.7
\]

Formally, restricting the sets \( \mathcal{V}_{\mathtt{GT}} \) and \( \mathcal{V}_{\mathtt{DB}} \) to values within a specific column \( \mathtt{col} \), CC is defined as:

\begin{equation}
    \mathrm{CC} = \frac{|\mathcal{V}_{\mathtt{GT}} \cap \mathcal{V}_{\mathtt{DB}}|}{|\mathcal{V}_{\mathtt{GT}}|}, \quad \text{where } \mathcal{V}_{\mathtt{GT}}, \mathcal{V}_{\mathtt{DB}} \in \mathtt{col}
\end{equation}

Here, the intersection counts only those values matched under semantically correct columns according to the cosine similarity criterion above.

\section{Experimentation Details}
\label{appendix:experiments}
\subsection{Experimental Setting}
\textbf{Computational Resources and Model Sizes.} We report the number of parameters, computational budget, and infrastructure details for all models and experiments used in this work. The models employed include: \textsc{LLaMA3-8B-Instruct} (8B parameters), \textsc{Claude 3.7 Sonnet} (parameter size not disclosed), \textsc{GPT-4o} (parameter size not disclosed), \textsc{Qwen3-8B} (8B parameters), and \textsc{DeepSeek-V2.5} (16B parameters). All experiments, including both development and final evaluation runs, were conducted using 1 GPU (NVIDIA A10, 24 GB VRAM) over a total of approximately 100 GPU hours. Our computing environment included 48-core Intel Xeon Silver 4310 CPUs and 128 GB RAM, running on Ubuntu 24.04.2 LTS. These details are provided to support reproducibility and contextualize the performance reported in this study.

\subsection{Results}
We report additional results from our study in this section. Table~\ref{tab:schema-coverage-domain} presents schema coverage scores across different domains and datasets for the \textsc{DeepSeek-v2.5} model and CoT approach. Table~\ref{tab:db_coverage_scores_all} shows the impact of different value sources on Tuple Coverage (\texttt{TC}), Value Coverage (\texttt{VC}) and Column Consistency (\texttt{CC}).
\vspace{2em}
\begin{table}[H]
\centering

\begin{tabular}{llccc}
\toprule
 & \textbf{Domain} & \textbf{ECS} & \textbf{PKC} & \textbf{FKC} \\
\midrule
\multirow{3}{*}{\rotatebox[origin=c]{90}{Kaggle}} 
  & Tourism              & 89.15 & 100 & 100 \\
  & Education            & \textbf{91.08} & 100 & 100 \\
  & Finance              & 84.71 & 100 & 100 \\
\midrule
\multirow{7}{*}{\rotatebox[origin=c]{90}{BIRD}} 
  & California Schools   & 76.84 & 100 & 100 \\
  & Superhero            & 77.55 & 100 & 100 \\
  & Books                & 84.87 & 100 & 100 \\
  & Computer Student     & 57.46 & 100 & 100 \\
  & Mental Health Survey & \textbf{38.97} & 100 & 100 \\
  & Authors              & 86.44 & 100 & 100 \\
\bottomrule
\end{tabular}

\caption{Schema coverage scores across different domains and datasets for the \textsc{DeepSeek-v2.5} model and CoT approach.}
\label{tab:schema-coverage-domain}
\end{table}

\begin{table*}[]
    \centering
    \resizebox{\textwidth}{!}{%

\begin{tabular}{lrcccccccccccccccccc}
\toprule
\textbf{Model}                              & \textbf{Diff.}   & \multicolumn{3}{c}{\mOne}   & \multicolumn{3}{c}{\mTwo}  & \multicolumn{3}{c}{\mThree} & \multicolumn{3}{c}{\eOne} & \multicolumn{3}{c}{\eTwo} & \multicolumn{3}{c}{\eThree} \\ 
& & \multicolumn{3}{c}{\small{(1)}}   & \multicolumn{3}{c}{\small{(2)}}  & \multicolumn{3}{c}{\small{(3)}} & \multicolumn{3}{c}{\small{(1)+(2)}} & \multicolumn{3}{c}{\small{(1)+(3)}} & \multicolumn{3}{c}{\small{(1)+(2)+(3)}} \\ 
                                   &                             & \texttt{TC}          & \texttt{VC}          & \texttt{CC}          & \texttt{TC}           & \texttt{VC}          & \texttt{CC}          & \texttt{TC}           & \texttt{VC}           & \texttt{CC}           & \texttt{TC}      & \texttt{VC}      & \texttt{CC}      & \texttt{TC}       & \texttt{VC}       & \texttt{CC}       & \texttt{TC}         & \texttt{VC}         & \texttt{CC}        \\ \hline
\multirow{4}{*}{\textsc{Claude 3.7 Sonnet}} & Easy                        & 1.00        & 0.97        & 0.97        & 1.00         & 0.97        & 0.97        & 0.98         & 0.93         & 0.93         & 1.00    & 0.98    & 0.98    & 1.00     & 0.97     & 0.97     & 1.00       & 0.98       & 0.98      \\
                                   & Med                      & 0.85        & 0.68        & 0.64        & 0.93         & 0.74        & 0.70        & 0.93         & 0.74         & 0.70         & 0.97    & 0.77    & 0.72    & 0.97     & 0.77     & 0.72     & 0.98       & 0.78       & 0.74      \\
                                   & Hard                        & 0.82        & 0.51        & 0.33        & 0.94         & 0.58        & 0.37        & 0.88         & 0.51         & 0.33         & 0.96    & 0.60    & 0.40    & 0.97     & 0.60     & 0.39     & 1.00       & 0.63       & 0.41      \\
                                   & Avg.                     & 0.89        & 0.72        & 0.65        & 0.96         & 0.76        & 0.68        & 0.93         & 0.72         & 0.65         & 0.98    & 0.78    & 0.70    & 0.98     & 0.78     & 0.70     & 0.99       & 0.80       & 0.71      \\ \hline
\multirow{4}{*}{\textsc{DeepSeek-v2.5}}     & Easy                        & 0.99        & 0.92        & 0.92        & 0.99         & 0.94        & 0.94        & 1.00         & 0.92         & 0.91         & 1.00    & 0.96    & 0.96    & 1.00     & 0.95     & 0.95     & 1.00       & 0.96       & 0.96      \\
                                   & Med                      & 0.96        & 0.76        & 0.71        & 0.87         & 0.68        & 0.64        & 0.90         & 0.69         & 0.66         & 0.98    & 0.79    & 0.75    & 0.98     & 0.80     & 0.76     & 0.99       & 0.80       & 0.77      \\
                                   & Hard                        & 0.89        & 0.54        & 0.35        & 0.83         & 0.42        & 0.26        & 0.75         & 0.35         & 0.24         & 0.95    & 0.57    & 0.38    & 0.95     & 0.57     & 0.38     & 0.95       & 0.59       & 0.39      \\
                                   & Avg.                     & 0.95        & 0.74        & 0.66        & 0.90         & 0.68        & 0.62        & 0.88         & 0.65         & 0.60         & 0.98    & 0.78    & 0.70    & 0.98     & 0.78     & 0.70     & 0.98       & 0.79       & 0.71      \\ \hline
\multirow{4}{*}{\textsc{GPT-4o}}            & Easy                        & 0.98        & 0.90        & 0.90        & 0.98         & 0.93        & 0.93        & 1.00         & 0.90         & 0.89         & 0.99    & 0.95    & 0.95    & 1.00     & 0.96     & 0.96     & 1.00       & 0.97       & 0.97      \\
                                   & Med                      & 0.94        & 0.75        & 0.71        & 0.88         & 0.69        & 0.66        & 0.90         & 0.68         & 0.64         & 0.98    & 0.80    & 0.76    & 0.98     & 0.81     & 0.77     & 0.99       & 0.81       & 0.77      \\
                                   & Hard                        & 0.81        & 0.50        & 0.32        & 0.78         & 0.41        & 0.29        & 0.93         & 0.51         & 0.33         & 0.92    & 0.56    & 0.38    & 0.97     & 0.59     & 0.38     & 0.97       & 0.61       & 0.40      \\
                                   & Avg.                     & 0.91        & 0.71        & 0.64        & 0.88         & 0.67        & 0.63        & 0.94         & 0.69         & 0.62         & 0.97    & 0.77    & 0.70    & 0.98     & 0.79     & 0.70     & 0.99       & 0.80       & 0.71      \\ \hline
\multirow{4}{*}{\textsc{Llama3-8B-Instruct}} & Easy                        & 0.99        & 0.89        & 0.88        & 0.81         & 0.74        & 0.74        & 0.77         & 0.61         & 0.60         & 1.00    & 0.95    & 0.95    & 1.00     & 0.94     & 0.94     & 1.00       & 0.95       & 0.95      \\
                                   & Med                      & 0.95        & 0.70        & 0.66        & 0.76         & 0.52        & 0.48        & 0.89         & 0.61         & 0.57         & 0.98    & 0.75    & 0.72    & 0.98     & 0.76     & 0.73     & 0.99       & 0.79       & 0.75      \\
                                   & Hard                        & 0.95        & 0.60        & 0.39        & 0.88         & 0.37        & 0.26        & 0.71         & 0.40         & 0.25         & 1.00    & 0.64    & 0.41    & 1.00     & 0.68     & 0.42     & 1.00       & 0.70       & 0.44      \\
                                   & Avg.                     & 0.96        & 0.73        & 0.64        & 0.81         & 0.54        & 0.49        & 0.79         & 0.54         & 0.47         & 0.99    & 0.78    & 0.69    & 0.99     & 0.79     & 0.70     & 1.00       & 0.81       & 0.71      \\ \hline
\multirow{4}{*}{\textsc{Qwen3-8B}}          & Easy                        & 0.99        & 0.92        & 0.92        & 0.97         & 0.92        & 0.92        & 0.85         & 0.72         & 0.72         & 1.00    & 0.96    & 0.96    & 1.00     & 0.96     & 0.96     & 1.00       & 0.96       & 0.96      \\
                                   & Med                      & 0.94        & 0.71        & 0.71        & 0.90         & 0.71        & 0.71        & 0.96         & 0.67         & 0.67         & 0.95    & 0.74    & 0.73    & 0.98     & 0.78     & 0.78     & 0.98       & 0.79       & 0.79      \\
                                   & Hard                        & 0.57        & 0.29        & 0.29        & 0.36         & 0.23        & 0.23        & 0.76         & 0.35         & 0.35         & 0.59    & 0.33    & 0.33    & 0.99     & 0.48     & 0.48     & 0.99       & 0.51       & 0.51      \\
                                   & Avg.                     & 0.83        & 0.64        & 0.64        & 0.75         & 0.62        & 0.62        & 0.85         & 0.58         & 0.58         & 0.85    & 0.68    & 0.67    & 0.99     & 0.74     & 0.74     & 0.99       & 0.76       & 0.75      \\ \hline
\bottomrule
\end{tabular}
    }
    \caption{Impact of different value sources on Tuple Coverage (\texttt{TC}), Value Coverage (\texttt{VC}) and Column Consistency (\texttt{CC}). \eOne combines tuples generated from \mOne and \mTwo while \eTwo combines \mOne and \mThree. \systemname~combines outputs from all three prompts. }
    \label{tab:db_coverage_scores_all}
\end{table*}

\subsection{Additional Context}
\textbf{Baseline Join Query vs \systemname Join Query.}
For the baseline case, the model was prompted to generate join queries after seeing the full table contents, allowing it to tailor joins to observed values. In contrast, \systemname’s join queries are issued independently of table population, which may result in more \texttt{None} retrievals.

\section{Related Work}
\label{appendix:related-work}

Recent research relevant to our task of synthesizing relational databases from unstructured text spans three primary areas: (1) summarizing structured information from text (2) interacting with or modifying existing databases (3) domain-specific, non-LLM approaches based on rule-based or statistical methods for relational structure extraction from text.

\paragraph{Summarizing Structures from Text.}  
A widely studied area related to our task is \textit{Open Information Extraction} (OpenIE), which extracts subject, predicate, object (SPO) triplets from unstructured text \cite{niklaus2018survey}. While OpenIE provides useful abstractions, the extracted triplets are not organized under a formal data model. A more structured alternative is the \textit{text-to-table} generation task. Early works approach this as a sequence modeling problem, jointly generating column headers and cell contents \cite{wu2022texttotable}. More recent systems such as \textbf{gTBLS}\cite{sundar2024gtbls} and \textbf{Seq2Seq\&Set}\cite{li2023seq2seqset} decouple schema inference from data population in the \textit{text-to-table} task, yielding improvements in table validity and structure. Other lines of work explore extracting structured data from semi-structured documents such as HTML and PDF using LLMs \cite{Aroravldb}, or schema-driven information extraction from heterogeneous tables \cite{bai2023schemadri}. However, the outputs remain flat and lack the normalized relationships central to relational database design. \textbf{T3}~\cite{deng2024texttupletableinformationintegrationtexttotable} takes a step further by converting extracted tuples into flat tables, which is conceptually closest to our use of intermediate triplet representations. Still, their method does not capture inter-table relationships, limiting alignment with relational database requirements. Additionally, other research explores non-relational structures such as mind maps for representing extracted information \cite{jain2024structsumgenerationfastertext}, which similarly do not align with the relational database model our work targets. 

\paragraph{Manipulating Existing Databases.}  
Another line of work focuses on interacting with or updating existing relational databases using language. Early work such as \cite{Mansuri-uns-to-db} proposed integrating unstructured sources into relational databases using information extraction and matching techniques, but relied heavily on statistical models, rule-based systems and domain-specific heuristics. In \textbf{TEXT2DB}~\cite{jiao-etal-2024-text2db}, LLM agents ingest documents and update a pre-existing relational database. While it operates on relational databases, it assumes an existing database with a predefined schema and does not attempt to synthesize a new one. On the other hand, the \textit{text-to-SQL} literature~\cite{hong2024next,yu2018spider} focuses on translating natural language queries into executable SQL statements over a known schema. Other works in this space include relation-aware schema encoding for better generalization \cite{pang2020ratsql}, constrained decoding for syntactically valid SQL generation \cite{scholak2021picard}, and synthetic data generation to improve model robustness \cite{chang2024synthesizingtexttosql}. However, none of these works attempt to synthesize a relational database from text.

\paragraph{Non-LLM Approaches for Relational Structure Extraction From Text.}  
Before LLMs, integrating unstructured text into relational databases relied on classical pipelines combining information extraction, schema induction, and entity linking. Systems such as \textbf{DeepDive}~\cite{zhang2016deepdive}, \textbf{LILLIE}~\cite{smith2022lillie}, and \textbf{OpenKI}~\cite{zhang2019openki} extracted structured facts and aligned them with relational schemas using statistical inference, symbolic reasoning, or context-aware matching. In web-centric domains, methods like \textbf{SEDE}~\cite{deng2010automatic,deng2011sede} and wrapper induction systems~\cite{carlson2008bootstrapping,chang2016fastwrapper,yuliana2016afis,yuliana2020dcade} inferred schemas from repeated HTML patterns and populated tables using DOM-based alignment. Statistical models such as \textbf{SICTF}~\cite{nimishakavi2016sictf} induced relation schemas from OpenIE triples via joint tensor factorization. These non-LLM methods demonstrated the feasibility of relational synthesis via symbolic or statistical reasoning, but typically required domain-specific tuning and struggled to generalize across diverse, noisy input text.

Broadly, existing research either aims to extract tables from text or to interface with predefined relational databases—without bridging the gap between the two. To our knowledge, no existing work performs fully automated and domain-generalized \textit{text-to-relational database synthesis}. Our system fills this gap by leveraging a neurosymbolic framework that decomposes the task into interpretable stages.

\section{Artifact Use}
\label{appendix:artifact}
\subsection{Dataset License Information}
In accordance with ACL guidelines, we disclose the licenses of all datasets used.

The BIRD benchmark datasets \cite{bird_benchmark} are distributed under various open licenses including Public Domain, CC0, CC-BY 4.0, CC-BY-SA 4.0, GPL, and CPOL, all permitting research use and redistribution.

The Kaggle datasets utilized in our experiments are licensed as follows, and all allow research use and redistribution:
\begin{itemize}
    \item \textbf{Tourism dataset} \cite{traveler_trip_2023}: Licensed under Creative Commons Attribution 4.0 (CC-BY 4.0).
    \item \textbf{Education dataset} \cite{education_kaggle}: Licensed under Creative Commons Attribution 4.0 (CC-BY 4.0).
    \item \textbf{Finance dataset} \cite{fintech_ltv_2023}: Licensed under the MIT License.
\end{itemize}

Additionally, we will release our generated dataset publicly under a \textsc{CC BY 4.0} License.

\subsection{Software and Language Models}
We used Stanford CoreNLP (v4.5.9) \cite{manning-etal-2014-stanford}, licensed under GNU GPLv3, which permits free use, modification, and redistribution under open-source terms.

The language models employed are publicly available and used under their respective license or terms of service:
\begin{itemize}
    \item \textbf{\textsc{LLaMa-3-8B-Instruct}} \cite{meta2024llama3}: Released under Meta’s research license allowing academic use.
    \item \textbf{\textsc{Claude 3.7 Sonnet}} \cite{anthropic2024claude3}: Provided under Anthropic’s terms for research and commercial use.
    \item \textbf{\textsc{GPT-4o}} \cite{openai2024gpt4o}: Accessed via OpenAI’s API under their usage policies.
    \item \textbf{\textsc{Qwen3-8B}} \cite{alibaba2024qwen3b}: Released with a permissive license for research use.
    \item \textbf{\textsc{DeepSeek-V2.5}} \cite{deepseek2024v25}: Licensed for research use as specified by DeepSeek AI.
\end{itemize}

\section{Prompts}
\label{appendix:prompts}
All of the prompts we use in \systemname are provided in Figures~\ref{appendix:prompt-dataset} to~\ref{appendix:baseline-prompts}.

\begin{figure*}[!t]
\begin{lstlisting}[basicstyle=\ttfamily\small, breaklines=true, frame=single, numbers=none, breakindent=0pt]
You are a creative AI that rephrases given sentences into engaging, conversational stories while incorporating all provided datapoints. 

- Ensure that no information is omitted or added, and skip any datapoints labeled as 'nan'. 
- Do not rephrase the object of a sentence. For example, if the sentence is 'start date is $9/22/2023$', do not change the date to a different format. 
- Respond only with the rephrased sentence without any additional commentary.
\end{lstlisting}
\caption{Prompts for dataset generation with \textsc{LLaMA3-8B-Instruct}: system prompt}
\label{appendix:prompt-dataset}
\end{figure*}


\begin{figure*}[t]
\begin{lstlisting}[basicstyle=\ttfamily\small, breaklines=true, frame=single, numbers=none, breakindent=0pt]
Rephrase the following sentence into a conversational story, ensuring all data points are included while skipping 'nan' values.
Do not introduce any extra or false details.

Original sentence: {sentence}

Creative sentence:
\end{lstlisting}
\caption{Prompts for dataset generation with \textsc{LLaMA3-8B-Instruct}: user prompt template}
\end{figure*}

\begin{figure*}[t]

\begin{lstlisting}[basicstyle=\ttfamily\small, breaklines=true, frame=single, numbers=none, breakindent=0pt]
You are an expert at formulating database schemas from textual data. I have given you a paragraph of text. 

Using this text, your task is to generate a relational database schema in JSON format. 

### **Task:**
1. **Extract Entities & Relationships**: Identify unique entity types and relationships.
2. **Determine Attributes**: Define necessary columns for each table.
3. **Normalize the Schema**: Ensure proper **primary keys, foreign keys, and normalization (3NF)**.
4. **Generate Output in JSON Format**.
5. **Column and Table name restriction**:

reserved_sql_keywords = ["order", "group", "select", "from", "where", "join", "on", "as", "and", "or", "by", "insert", "update", "delete", "create", "drop", "alter", "into", "table"]
- Ensure that the table names and column names do not contain any SQL reserved keywords. 
\end{lstlisting}
\caption{Prompts for schema generation: system prompt}
\label{appendix:prompt-schema-direct}
\end{figure*}

\begin{figure*}[t]
\begin{lstlisting}[basicstyle=\ttfamily\small, breaklines=true, frame=single, numbers=none, breakindent=0pt]
### **Text:**
{text}

### **Expected Example Output Format (Strictly Follow This Structure while modifying the table_names, column_names to match the given text)**:
schema = [
    {{
        "table_name": "student",
        "columns": [
            {{"name": "id", "type": "INTEGER", "primary_key": True}},
            {{"name": "name", "type": "TEXT"}},
        ]
    }},
    {{
        "table_name": "course",
        "columns": [
            {{"name": "id", "type": "INTEGER", "primary_key": True}},
            {{"name": "title", "type": "TEXT"}},
        ]
    }},
    {{
        "table_name": "enrollment",
        "columns": [
            {{"name": "id", "type": "INTEGER", "primary_key": True}},
            {{"name": "student_id", "type": "INTEGER", "foreign_key": True, "foreign_key_table": "student", "foreign_key_column": "id"}},
            {{"name": "course_id", "type": "INTEGER", "foreign_key": True, "foreign_key_table": "course", "foreign_key_column": "id"}}
        ]
    }}
]


Now output the schema as per the system instructions.
### Output:
\end{lstlisting}
\caption{Prompts for schema generation: user prompt template}
\end{figure*}

\label{appendix:prompt-schema-cot}
\begin{figure*}[t]
\begin{lstlisting}[basicstyle=\ttfamily\small, breaklines=true, frame=single, numbers=none, breakindent=0pt]
You are an expert at formulating database schemas from textual data. I have given you a paragraph of text. 
Using this text, your task is to generate a relational database schema in JSON format.   

---

### **Step-by-Step Guide for Schema Creation (Follow This Chain of Thought)**  

**Requirements Analysis**:  
- Identify all distinct entities and attributes from the text.  
- Determine necessary tables and their columns.  

**Entity-Relationship (ER) Modeling**:  
- Identify entity relationships (One-to-One, One-to-Many, Many-to-Many).  
- If applicable, use associative tables for Many-to-Many relationships.  

**Define Tables and Columns**:  
- Convert entities into relational tables with appropriate **data types**.  

**Establish Primary Keys**:  
- Assign a **Primary Key (PK)** for each table to uniquely identify records.  

**Define Relationships and Foreign Keys**:  
- Use **Foreign Keys (FK)** to enforce referential integrity between tables.  
- Ensure that it is possible to join all tables to create one flat table using the foreign keys.
- Apply **ON DELETE CASCADE** if necessary to maintain consistency.  

**Normalization (1NF => 2NF => 3NF)**:  
- Ensure atomic values (1NF).  
- Remove partial dependencies (2NF).  
- Eliminate transitive dependencies (3NF).  

**Define Constraints**:  
- Apply **NOT NULL**, **UNIQUE**, **CHECK**, and other constraints as needed.  

**Indexing for Performance**:  
- Create indexes on frequently queried columns (e.g., search fields).   

- **Column and Table name restriction**:
reserved_sql_keywords = ["order", "group", "select", "from", "where", "join", "on", "as", "and", "or",
        "by", "insert", "update", "delete", "create", "drop", "alter", "into", "table"]
- Ensure that the table names and column names do not contain any SQL reserved keywords. 

---
### **Task Instructions:**  
- **Step through the schema creation process using the above guide**.  
- **Generate a well-structured, normalized relational database schema**.  
- **Output only the final schema** in Python dictionary format (NO explanations).
- **Column and Table name restriction**:
reserved_sql_keywords = ["order", "group", "select", "from", "where", "join", "on", "as", "and", "or",
        "by", "insert", "update", "delete", "create", "drop", "alter", "into", "table"]
- Ensure that the table names and column names do not contain any SQL reserved keywords. 

\end{lstlisting}
\caption{Prompts for schema generation: CoT - system prompt}
\end{figure*}

\begin{figure*}[t]
\begin{lstlisting}[basicstyle=\ttfamily\small, breaklines=true, frame=single, numbers=none, breakindent=0pt]
### **Text:**
{text}

### **Expected Example Output Format (Strictly Follow This Structure while modifying the table_names, column_names to match the given text)**:
schema = [
    {{
        "table_name": "student",
        "columns": [
            {{"name": "id", "type": "INTEGER", "primary_key": True}},
            {{"name": "name", "type": "TEXT"}},
        ]
    }},
    {{
        "table_name": "course",
        "columns": [
            {{"name": "id", "type": "INTEGER", "primary_key": True}},
            {{"name": "title", "type": "TEXT"}},
        ]
    }},
    {{
        "table_name": "enrollment",
        "columns": [
            {{"name": "id", "type": "INTEGER", "primary_key": True}},
            {{"name": "student_id", "type": "INTEGER", "foreign_key": True, "foreign_key_table": "student", "foreign_key_column": "id"}},
            {{"name": "course_id", "type": "INTEGER", "foreign_key": True, "foreign_key_table": "course", "foreign_key_column": "id"}}
        ]
    }}
]

Now output the schema as per the system instructions.
### Output:
\end{lstlisting}
\caption{Prompts for schema generation: CoT - user prompt template}
\end{figure*}

\begin{figure*}[t]
\label{appendix:prompt-triplet}
\begin{lstlisting}[basicstyle=\ttfamily\small, breaklines=true, frame=single, numbers=none, breakindent=0pt]
You are a helpful assistant that who assists a user with information extraction tasks. 
Your job is to associate a unique superkey value with each paragraph in the text. 
You will be given multiple paragraphs of text, a database schema, and a superkey. 
Your task is to associate the superkey value with each paragraph in the text. Each paragraph MUST be associated with a superkey value. No two superkey values should be the same.
Fill in the <FILL IN WITH APPROPRIATE VALUE OF {superkey}> with the value. You will not provide code or SQL, you will do the task yourself.
\end{lstlisting}
\caption{Prompts for triplet generation with LLM- unique identifier association: system prompt}
\end{figure*}

\begin{figure*}[t]
\begin{lstlisting}[basicstyle=\ttfamily\small, breaklines=true, frame=single, numbers=none, breakindent=0pt]
**Text**:
{text}

**Schema**:
{schema}

**Superkey**:
{superkey}

**Paragraphs**:
"""
    user_prompt += "\n--\n"
    for j in range(len(paragraphs)):
        user_prompt += f"paragraph {j}: {paragraphs[j]}\n"
        user_prompt += f"associated superkey: <FILL IN WITH APPROPRIATE VALUE OF {superkey}>\n"
        user_prompt += "\n--\n"
\end{lstlisting}
\caption{Prompts for triplet generation with LLM- unique identifier association: user prompt template}
\end{figure*}

\begin{figure*}[t]
\begin{lstlisting}[basicstyle=\ttfamily\small, breaklines=true, frame=single, numbers=none, breakindent=0pt]
You are an expert in Open Information Extraction and relational databases. Given a database schema and a natural language paragraph, your task is to extract all factual information from each sentence of the paragraph in the form of triplets, structured as a Python list of dictionaries. 

Each dictionary should have the following keys: 'table_name', 'column_name', and 'value'. Ensure that the extracted triplets strictly follow the format: {'table_name': <table_name>, 'column_name': <column_name>, 'value': }. 

Only extract values that explicitly appear in the input sentence. The table_name and column_name must match the schema. Do not invent values or infer unstated facts. You don’t need to generate triplets for values that are not mentioned. DO NOT generate code, do the task yourself.
\end{lstlisting}
\caption{Prompts for triplet generation with LLM- triplet generation: system prompt}
\label{appendix:LLM-gen-triplet-prompt}
\end{figure*}

\begin{figure*}[t]
\begin{lstlisting}[basicstyle=\ttfamily\small, breaklines=true, frame=single, numbers=none, breakindent=0pt]
You will be given a database schema and a sentence. Extract all relevant triplets of the form:
{{"table_name": <table_name>, "column_name": <column_name>, "value": <value>}}
Your output must be a valid Python list of dictionaries. Do not include any explanations or notes-only return the list.
{example_schema}

Sentence: {example_text}

{example_output}

Now extract triplets for the following input:

Schema:
{schema}

Sentence: {text}
Triplets:

\end{lstlisting}
\caption{Prompts for triplet generation with LLM- triplet generation: user prompt template}
\end{figure*}

\begin{figure*}[t]
\label{appendix:prompt-value pop}
\begin{lstlisting}[basicstyle=\ttfamily\small, breaklines=true, frame=single, numbers=none, breakindent=0pt]
Extract information using the extraction tool -- "extract".

Example:
Python table schema: 
    - person: "id: int [PK]", "name: string", "age: int", "location: string", "married: boolean"
    - job: "id: int [PK]", "person_id: int [FK => person(id)]", "title: string", "salary: int", "department: string"


Example triplets:
    < 1 , person, name, John >
    < 1 , person, age, 30 >
    < 1 , person, location, Los Angeles >
    < 1 , person, married, false >
    < 1 , job, title, Software Engineer >
    < 1 , job, salary, 100000 >
    < 1 , job, department, Engineering >
    
Example usage: 
    extract person: "id": 1; "name": "John"; "age" 30; "location": "Los Angeles"; "married": false
    extract job: "id": 1; "person_id": 1; "title": "Software Engineer"; "salary": 100000; "department": "Engineering"

Tool guidelines:
- "id" should always be present in the extraction. It is the primary key of the table. You must assign an appropriate value to it. Do not leave it empty.
- "xxx_id" is the foreign key of the table and references the primary key of a row in another table. It must also be present in the extraction, you must assign an appropriate value to it, and ensure that the value matches the id of the row in it's foreign key parent table. Do not leave it empty.
- You must follow the order of the columns per table, defined in the Python table schema.
- You must follow the data type for each column.
- You must use "extract" once per table of the schema to extract information from the whole paragraph and the set of triplets for each of the table.
- After reading the paragraph, the schema and the triplets, do all the extraction steps in one go, do not skip any extraction steps. Do not repeat extraction for the same id if there are no new information.
- Do not extract any information that is not present in the original paragraph.
- Other than "id" and "xxx_id", if no value is found for a column, use '?' to represent the value. If all columns are empty, do not use extract for that row.

Python table schema:
{table_instruction_str}
\end{lstlisting}
\caption{Prompts for table population: tooluse prompt for extraction}
\end{figure*}

\begin{figure*}[t]
\begin{lstlisting}[basicstyle=\ttfamily\small, breaklines=true, frame=single, numbers=none, breakindent=0pt]
system_prompt = f"""You are an expert at populating values in a database from text based on a given database schema. I have provided a paragraph of text and a database schema. Using this information, your task is to extract relevant values and format them according to the schema. Do not provide code, do the task yourself."""
    user_prompt = f"""
### **Text:**
{text}
### **Schema:**
{schema}
### **Expected Output Format:**
{data_template}
Output as many rows as necessary to populate the data. Replace the '#' with the actual values from the text.

You will follow this chain-of-thought reasoning to generate the final output:
- Generate output entries relevant to the text.
- Follow the given output format strictly. Do not add any additional explanations or comments. Only output the data entries in given format. Do not provide code, do the task yourself.
### **Output:**
"""
\end{lstlisting}
\caption{Prompts for table population: Method \mOne}
\end{figure*}

\begin{figure*}[t]
\begin{lstlisting}[basicstyle=\ttfamily\small, breaklines=true, frame=single, numbers=none, breakindent=0pt]
text = data['text']
schema = data['schema']
superkey = data['superkey']
data_template = generate_empty_data_template_tooluse(schema)
identified_values = data['identified_values']

system_prompt = f"""You are a database assistant that extracts data from each sentence of a given text, and populates data entries into a fixed relational database schema.

The user will give you the following inputs: a text with data of multiple users, the primary identifier (superkey) of that text, extracted triplets from that text in the following format:
<superkey, `subject`, `relation`, `object`>

You will follow this chain-of-thought reasoning to generate the final output:
- Generate output entries as per given instructions, relevant to the current paragraph.
- Validate each output entry by going through the triplets one by one, and ensure every unique data point from the triplets is captured into the correct table and fields.
- Follow the given output format strictly. Do not add any additional explanations or comments. Only output the data entries in given format. Do not provide code, do the task yourself.

Text:
{text}

Output format:
{data_template}
\end{lstlisting}
\caption{Prompts for table population: Method \mTwo}
\end{figure*}

\begin{figure*}[t]
\begin{lstlisting}[basicstyle=\ttfamily\small, breaklines=true, frame=single, numbers=none, breakindent=0pt]
text = data['text']
schema = data['schema']
superkey = data['superkey']
data_template = generate_empty_data_template_tooluse(schema)
identified_values = data['identified_values']

system_prompt = f"""You are a database assistant that extracts data from each sentence of a given text, and populates data entries into a fixed relational database schema.

The user will give you the following inputs: a text with data of multiple users, the primary identifier (superkey) of that text, extracted triplets from that text in the following format:
<superkey, `table_name`, `column_name`, `value`>

You will follow this chain-of-thought reasoning to generate the final output:
- Generate output entries as per given instructions, relevant to the current paragraph.
- Validate each output entry by going through the triplets one by one, and ensure every unique data point from the triplets is captured into the correct table and fields.
- Follow the given output format strictly. Do not add any additional explanations or comments. Only output the data entries in given format. Do not provide code, do the task yourself.

Text:
{text}

Output format:
{data_template}

"""
    user_prompt = "Please perform the task as per the system instructions.\n"
    user_prompt += f"### Extracted <{superkey}, `table_name`, `column_name`, `value`> triplets from each paragraph:\n"
    for item in identified_values:
        for triplet in item['triplets']:
            if triplet['value'] !="not mentioned" or triplet['value'] !="not provided":
                user_prompt += f"<{item['superkey']}, {triplet['table_name']}, {triplet['column_name']}, {triplet['value']}>\n"
        user_prompt += "\n"
    user_prompt += "### Output:\n"
\end{lstlisting}
\caption{Prompts for table population: Method \mThree}
\end{figure*}

\begin{figure*}[t]

\begin{lstlisting}[basicstyle=\ttfamily\small, breaklines=true, frame=single, numbers=none, breakindent=0pt]
system_prompt = """You are a database expert. Your task is generating SQL 'CREATE TABLE' and 'INSERT INTO' statements from text."""
    user_prompt = f"""
### **Text:**
{text}
### **Output: (Write the create table and insert into statements together)
"""
\end{lstlisting}

\begin{lstlisting}[language=Python, basicstyle=\ttfamily\small, breaklines=true, frame=single, numbers=none]
system_prompt = """You are a database expert. Your task is generating a sqlite join query from 'create table' and 'insert into' statements."""
    user_prompt = f"""
### **'CREATE TABLE' and 'INSERT INTO' statements:**
{sql_statements}
### **Output:**
"""
\end{lstlisting}
\caption{Prompts for baseline}
\label{appendix:baseline-prompts}
\end{figure*}


\end{appendix}

\end{document}